\documentclass[11pt,fleqn]{article}
\usepackage{amssymb,latexsym,amsmath,amsfonts,graphicx}

    \numberwithin{equation}{section}

    \def\ds{\displaystyle}
    
    \def\Re{{\rm Re \,}}
    \def\Im{{\rm Im \,}}
    \def\Ai{{\rm Ai \,}}
    \def\I{{\rm I \,}}
    \def\II{{\rm II \,}}
    \def\III{{\rm III \,}}
    \def\IV{{\rm IV \,}}
    \def\bigO{{\cal O}}

    \newtheorem{theorem}{Theorem}[section]
    
    \newtheorem{corollary}[theorem]{Corollary}
    \newtheorem{proposition}[theorem]{Proposition}
    
    \newtheorem{Definition}[theorem]{Definition}
    
    \newtheorem{Remark}[theorem]{Remark}
    \newenvironment{remark}{\begin{Remark}\rm}{\end{Remark}}
    \newtheorem{Example}[theorem]{Example}

    \newenvironment{proof}%
    {\rm \trivlist \item[\hskip \labelsep{\bf Proof. }]}%
    {\hspace*{\fill}$\Box$\endtrivlist}
    \newenvironment{varproof}%
    {\rm \trivlist \item[\hskip \labelsep{\bf Proof}]}%
    {\hspace*{\fill}$\Box$\endtrivlist}

    \newcommand{\supp}{{\operatorname{supp}}}
    \newcommand{\sgn}{{\operatorname{sgn}}}
    
    \DeclareMathOperator*{\Tr}{Tr}

\begin{document}

\title{Multi-critical unitary random matrix ensembles\\
and the general Painlev\'e II equation}
\author{T. Claeys, A.B.J. Kuijlaars, and M. Vanlessen}
\date{\ }
\maketitle

\begin{abstract} We study unitary random matrix ensembles of the form
$Z_{n,N}^{-1} |\det M|^{2\alpha} e^{-N \Tr V(M)}dM$, where $\alpha>-1/2$ and $V$
is such that the limiting mean eigenvalue density for $n,N\to\infty$ and $n/N\to 1$
vanishes quadratically at the origin. In order to compute the
double scaling limits of the eigenvalue correlation kernel near
the origin, we use the Deift/Zhou steepest descent
method applied to the Riemann-Hilbert problem for orthogonal
polynomials on the real line with respect to the weight
$|x|^{2\alpha}e^{-NV(x)}$.
Here the main focus is on the construction of a
local parametrix near the origin with $\psi$-functions associated with a special
solution $q_\alpha$ of the Painlev\'e II equation $q''=sq+2q^3-\alpha$.
We show that $q_\alpha$ has no real poles for $\alpha > -1/2$, by
proving the solvability of the corresponding Riemann-Hilbert
problem.
We also show that the asymptotics of the recurrence coefficients of the
orthogonal polynomials can be expressed in terms of $q_\alpha$ in the double scaling limit.
\end{abstract}

\section{Introduction and statement of results}

\subsection{Unitary random matrix ensembles}

For $n \in \mathbb N$, $N > 0$, and $\alpha > -1/2$, we consider the unitary random matrix
ensemble
\begin{equation} \label{1-randommatrixmodel}
    Z_{n,N}^{-1} |\det M|^{2\alpha} e^{-N \Tr V(M)} \;dM,
\end{equation}
on the space of $n\times n$ Hermitian matrices $M$, where $V : \mathbb R \to \mathbb R$ is a
real analytic function satisfying
\begin{equation} \label{1-voorwaardeV}
    \lim_{x\to\pm\infty} \frac{V(x)}{\log(x^2+1)} =
        +\infty.
\end{equation}
Because of (\ref{1-voorwaardeV}) and $\alpha > -1/2$, the integral
\begin{equation} \label{1-partition}
     Z_{n,N} =
        \int|\det M|^{2\alpha} e^{-N \Tr V(M)} \;dM
\end{equation}
converges and the matrix ensemble (\ref{1-randommatrixmodel}) is well-defined. It is well
known, see for example \cite{Deift,Mehta}, that the eigenvalues of $M$ are distributed
according to a determinantal point process with a correlation kernel given by
\begin{equation} \label{1-kern}
    K_{n,N}(x,y) =
        |x|^{\alpha}  e^{-\frac{N}{2}V(x)} |y|^{\alpha} e^{-\frac{N}{2}V(y)}
        \sum_{k=0}^{n-1} p_{k,N}(x) p_{k,N}(y),
\end{equation}
where $p_{k,N} = \kappa_{k,N} x^k + \cdots$, $\kappa_{k,N} > 0$, denotes the $k$-th degree
orthonormal polynomial with respect to the weight $|x|^{2\alpha} e^{-NV(x)}$ on $\mathbb R$.

Scaling limits of the kernel (\ref{1-kern}) as $n, N \to \infty$, $n/N \to 1$, show a
remarkable universal behavior which is determined to a large extent by the limiting mean
density of eigenvalues
\begin{equation} \label{1-limitingdensity}
    \psi_V(x) =
        \lim_{n \to \infty} \frac{1}{n} K_{n,n}(x,x).
\end{equation}
Indeed, for the case $\alpha = 0$, Bleher and Its \cite{BI1} (for
quartic $V$) and Deift et al.\ \cite{DKMVZ2} (for general real
analytic $V$) showed that the sine kernel is universal in the bulk
of the spectrum, i.e.,
\[
    \lim_{n \to \infty} \frac{1}{n \psi_V(x_0)}
    K_{n,n}\left(x_0+\frac{u}{n \psi_V(x_0)},x_0+\frac{v}{n \psi_V(x_0)}\right) =
        \frac{\sin \pi (u-v)}{\pi(u-v)}
\]
whenever $\psi_V(x_0) > 0$. In addition, the Airy kernel appears
generically at endpoints of the spectrum. If $x_0$ is a right
endpoint and $\psi_V(x) \sim (x_0-x)^{1/2}$ as $x \to x_0-$, then
there exists a constant $c>0$ such that
\[
    \lim_{n \to \infty} \frac{1}{cn^{2/3}}
    K_{n,n}\left(x_0 + \frac{u}{c n^{2/3}}, x_0 + \frac{v}{c n^{2/3}}\right) =
        \frac{\Ai(u) \Ai'(v) - \Ai'(u) \Ai(v)}{u-v},
\]
where $\Ai$ denotes the Airy function, see also \cite{DG2}.

The extra factor $|\det M|^{2\alpha}$ in
(\ref{1-randommatrixmodel}) introduces singular behavior at $0$ if
$\alpha \neq 0$. The pointwise limit (\ref{1-limitingdensity})
does not hold if $\psi_V(0) > 0$,  since $K_{n,n}(0,0) = 0$ if
$\alpha > 0$ and $K_{n,n}(0,0) = +\infty$ if $\alpha < 0$, due to
the factor $|x|^{\alpha} |y|^{\alpha}$ in (\ref{1-kern}). However
(\ref{1-limitingdensity}) continues to hold for $x \neq 0$ and
also in the sense of weak$^*$ convergence of probability measures
\[
    \frac{1}{n} K_{n,n}(x,x)dx \stackrel{*}{\to} \psi_V(x)dx,
        \qquad \mbox{as $n \to \infty$.}
\]
Therefore we can still call $\psi_V$ the limiting mean density of eigenvalues.
Observe that $\psi_V$ does not depend on $\alpha$.

However, at a microscopic level the introduction of the factor
$|\det M|^{2\alpha}$ changes the eigenvalue correlations near the
origin. Indeed, for the case of a non-critical $V$ for which
$\psi_V(0)>0$, it was shown in \cite{KV2} that
\begin{multline}
    \lim_{n\to\infty} \frac{1}{n\psi_V(0)}
    K_{n,n}\left(\frac{u}{n\psi_V(0)},\frac{v}{n\psi_V(0)}\right) =
    \\[1ex]
            \pi\sqrt{u}\sqrt{v}
            \frac{J_{\alpha+\frac{1}{2}}(\pi u)J_{\alpha-\frac{1}{2}}(\pi v)-
            J_{\alpha-\frac{1}{2}}(\pi u)J_{\alpha+\frac{1}{2}}(\pi v)}{2(u-v)},
\end{multline}
where $J_{\nu}$ denotes the usual Bessel function of order $\nu$.

We notice that universality results for orthogonal and symplectic
ensembles of random matrices have been obtained only very
recently, see \cite{DG1,DG2,DGKV}.

\subsection{Multi-critical case}

It is the goal of this paper to study (\ref{1-randommatrixmodel})
in a critical case where $\psi_V$ vanishes quadratically at $0$,
i.e.,
\begin{equation} \label{singularII}
    \psi_V(0)=\psi_V'(0)=0, \qquad \mbox{and} \qquad \psi_V''(0)>0.
\end{equation}
The behavior (\ref{singularII}) is among the possible singular
behaviors that were classified in \cite{DKM}.
The classification depends on the characterization of the
measure $\psi_V(x) dx$ as the unique minimizer of
the logarithmic energy
\begin{equation} \label{defIV}
    I_V(\mu)=\iint\log\frac{1}{|x-y|}d\mu(x)d\mu(y) +\int V(x)d\mu(x)
\end{equation}
among all probability measures $\mu$ on $\mathbb R$. The corresponding
Euler-Lagrange variational conditions give that for some constant
$\ell\in\mathbb{R}$,
\begin{align}
    \label{variational condition 1-psiV}
    & 2\int\log|x-y|\psi_V(y)dy-V(x)+\ell=0, && \mbox{for $x\in \supp(\psi_V)$,}
    \\
    \label{variational condition 2-psiV}
    & 2\int\log|x-y|\psi_V(y)dy-V(x)+\ell\leq 0, && \mbox{for $x \in \mathbb R$.}
\end{align}
In addition one has that $\psi_V$ is supported on a finite union
of disjoint intervals, and
\begin{equation} \label{qVminus}
    \psi_V(x) = \frac{1}{\pi} \sqrt{Q_V^-(x)},
    \end{equation}
where $Q_V$ is a real analytic function, and $Q_V^-$ denotes its
negative part. Note that the endpoints of the support correspond
to zeros of $Q_V$ with odd multiplicity.

The possible singular behaviors are as follows, see \cite{DKM,KM}.
\begin{description}
    \item[Singular case I] $\psi_V$ vanishes at an endpoint to higher order than a square root.
        This corresponds to a zero of the function $Q_V$ in (\ref{qVminus})
        of odd multiplicity $4k+1$ with $k\geq 1$. (The multipicity $4k+3$
        cannot occur in these matrix models.)
    \item[Singular case II] $\psi_V$ vanishes at an interior point of $\supp(\psi_V)$,
        which corresponds to a zero of $Q_V$ in the interior of the
        support. The multiplicity of such a zero is necessarily a multiple of $4$.
    \item[Singular case III] Equality holds in the variational inequality
        (\ref{variational condition 2-psiV}) for some $x \in \mathbb R \setminus\supp(\psi_V)$.
\end{description}
In each of the above cases, $V$ is called singular, otherwise regular.
The above conditions correspond to a singular endpoint, a singular interior point,
and a singular exterior point, respectively.

In each of the singular cases one expects a family of possible limiting kernels
in a double scaling limit as $n,N \to \infty$ and $n/N \to 1$ at some critical
rate \cite{BE}. As said before we consider the case (\ref{singularII}) which
corresponds to the singular case II  with $k=1$ at the singular point $x=0$.
For technical reasons we assume that there are no other singular points besides
$0$. Setting $t = n/N$, and letting $n, N \to \infty$ such that $t \to 1$, we
have that  the parameter $t$ describes the transition from the case where
$\psi_V(0)>0$ (for $t>1$) through the multi-critical case ($t=1$) to the case
where $0$ lies in a gap between two intervals of the spectrum ($t<1$). The
appropriate double scaling limit will be such that the limit
$\lim_{n,N\to\infty} n^{2/3}\left(t-1\right)$ exists.

The double scaling limit for $\alpha = 0$ was considered in
\cite{BDJ,BI2,BI3} for certain special cases, and in \cite{CK} in
general. The limiting kernel is built out of $\psi$-functions
associated with the Hastings-McLeod solution \cite{HastingsMcLeod} of the Painlev\'e II
equation $q'' = sq + 2q^3$.

For general $\alpha > -1/2$, we are led to the general Painlev\'e II
equation
\begin{equation} \label{PIIvgl}
    q'' = sq + 2q^3 - \alpha.
\end{equation}
The Painlev\'e II equation for general $\alpha$ has been suggested by
the physical papers \cite{ADMN, SeibergShih}. The limiting kernels in the double
scaling limit are associated with a special distinguished solution
of (\ref{PIIvgl}), which we describe first. We assume from now on
that $\alpha \neq 0$.

\subsection{General Painlev\'e II equation}
Balancing $sq$ and $\alpha$  in the differential
equation (\ref{PIIvgl}), we find that there exist solutions  such that
\begin{equation} \label{asymptotics q +infty}
    q(s) \sim \frac{\alpha}{s}, \qquad
        \mbox{as $s\to +\infty$,}
\end{equation}
and balancing $sq$ and $2q^3$, we see that there also exist solutions of
(\ref{PIIvgl}) such that
\begin{equation} \label{asymptotics q -infty}
    q(s) \sim \sqrt{\frac{-s}{2}}, \qquad
        \mbox{as $s\to -\infty$.}
\end{equation}
There is exactly one solution of (\ref{PIIvgl}) that satisfies
both (\ref{asymptotics q +infty}) and (\ref{asymptotics q -infty}),
see \cite{IK1,IK2,Kap}, and we denote it by $q_{\alpha}$.
This is the special solution that we need. It corresponds
to the choice of Stokes multipliers
\[
    s_1 = e^{-\pi i\alpha}, \qquad
    s_2 = 0, \qquad
    s_3 = -e^{\pi i \alpha},
\]
see Section 2 below.
We call $q_{\alpha}$ the Hastings-McLeod solution of the general Painlev\'e II
equation (\ref{PIIvgl}), since it seems to be the natural analogue
of the Hastings-McLeod solution for $\alpha = 0$.

The Hastings-McLeod solution is meromorphic in $s$ (as are all
solutions of (\ref{PIIvgl})) with an infinite number of poles. We
need that it has no poles on the real line. From the asymptotic
behavior (\ref{asymptotics q +infty}) and (\ref{asymptotics q
-infty}) we know that there are no real poles for $|s|$ large
enough, but that does not exclude the possibility of a finite
number of real poles. While there is a substantial literature on
Painlev\'e equations and Painlev\'e transcendents, we have not
been able to find the following result.

\begin{theorem}\label{theorem: polen q}
    Let $q_{\alpha}$ be the Hastings-McLeod solution of the general
    Painlev\'e II equation {\rm (\ref{PIIvgl})} with $\alpha > -1/2$.
    Then $q_{\alpha}$ is a meromorphic function with no poles on the real line.
\end{theorem}

\subsection{Main result}
To describe our main result, we recall the notion of $\psi$-functions
associated with the Painlev\'e II equation, see \cite{FlaschkaNewell}.
The Painlev\'e II equation (\ref{PIIvgl})
is the compatibility condition for the following system of linear
differential equations for $\Psi=\Psi_{\alpha}(\zeta;s)$.
\begin{equation}\label{systeemdiffvgln}
    \frac{\partial \Psi}{\partial \zeta}= A\Psi,
    \qquad \frac{\partial\Psi}{\partial s}= B\Psi,
\end{equation}
where
\begin{equation} \label{AenB}
    A =
        \begin{pmatrix}
            -4i\zeta^2-i(s+2q^2) & 4\zeta q+2ir+\alpha /\zeta  \\
            4\zeta q -2i r+\alpha /\zeta   & 4i\zeta^2+i(s+2q)
        \end{pmatrix},
    \qquad \mbox{and} \qquad
    B =
        \begin{pmatrix}
            -i\zeta & q \\
            q & i\zeta
        \end{pmatrix}.
\end{equation}
That is, (\ref{systeemdiffvgln}) has a solution where $q = q(s)$ and $r=r(s)$
depend on $s$ but not on $\zeta$, if and only if $q$ satisfies Painlev\'e II
and $r = q'$.

Given $s$, $q$ and $r$, the solutions of
\begin{equation} \label{equationsphi}
    \frac{\partial}{\partial \zeta}
    \begin{pmatrix}
        \Phi_1(\zeta) \\
        \Phi_2(\zeta)
    \end{pmatrix} =
            A
            \begin{pmatrix}
                \Phi_1(\zeta)\\
                \Phi_2(\zeta)
            \end{pmatrix}
\end{equation}
are analytic with branch point at $\zeta = 0$.
For $\alpha > -1/2$ and $s \in \mathbb R$, we take
$q = q_{\alpha}(s)$ and $r = q_{\alpha}'(s)$ where $q_{\alpha}$ is the Hastings-McLeod solution
of the Painlev\'e II equation, and we define
$\begin{pmatrix}\Phi_{\alpha,1}(\zeta;s) \\ \Phi_{\alpha,2}(\zeta;s) \end{pmatrix}$
as the unique solution of (\ref{equationsphi}) with asymptotics
\begin{equation}\label{asymptotiekphi}
    e^{i(\frac{4}{3}\zeta^3+s\zeta)}
    \begin{pmatrix}\Phi_{\alpha,1}(\zeta;s)\\
    \Phi_{\alpha,2}(\zeta;s)\end{pmatrix}
    =\begin{pmatrix}1\\0\end{pmatrix}+\bigO(\zeta^{-1}),
\end{equation}
uniformly as $\zeta\to\infty$ in the sector $ \varepsilon<\arg \zeta <\pi
-\varepsilon$ for any $\varepsilon>0$. Note that this is well-defined for every
$s \in \mathbb R$ because of Theorem \ref{theorem: polen q}.

The functions $\Phi_{\alpha,1}$ and $\Phi_{\alpha,2}$ extend to analytic
functions on $\mathbb C \setminus (-i\infty, 0]$, which we also denote
by $\Phi_{\alpha,1}$ and $\Phi_{\alpha,2}$, see also Remark \ref{definition: Phi_alpha}
below. Their values on the real line appear in the limiting kernel.
The following is the main result of this paper.

\begin{theorem}\label{theorem: main result}
    Let $V$ be real analytic on $\mathbb R$ such that {\rm
    (\ref{1-voorwaardeV})} holds. Suppose that $\psi_V$ vanishes
    quadratically in the origin, i.e., $\psi_V(0) = \psi_V'(0) = 0$,
    and $\psi_V''(0) > 0$, and that there are no other singular points
    besides $0$. Let $n, N \to \infty$ such that
    \[
        \lim_{n,N \to \infty} n^{2/3} (n/N - 1) = L \in \mathbb R
    \]
    exists.  Define constants
    \begin{equation} \label{definition: c}
        c = \left(\frac{\pi \psi_V''(0)}{8}\right)^{1/3},
    \end{equation}
    and
    \begin{equation} \label{definition: s}
        s = 2 \pi^{2/3} L \left[\psi_V''(0) \right]^{-1/3} w_{S_V}(0),
    \end{equation}
    where $w_{S_V}$ is the equilibrium density of the support of $\psi_V$
    (see Remark {\rm \ref{rem:equilibrium}} below). Then
    \begin{equation}\label{eqtheorem}
        \lim_{n,N\to\infty} \frac{1}{c n^{1/3}}
        K_{n,N}\left(\frac{u}{cn^{1/3}}, \frac{v}{cn^{1/3}}\right) =
                K^{crit,\alpha}(u,v;s),
    \end{equation}
    uniformly for $u, v$ in compact subsets of $\mathbb R\setminus\{0\}$, where
    \begin{align}\label{Kcrit}
        & K^{crit,\alpha}(u,v;s)= -e^{\frac{1}{2}\pi i\alpha[\sgn(u)+\sgn(v)]}
        \frac{\Phi_{\alpha,1}(u;s) \Phi_{\alpha,2}(v;s) -
        \Phi_{\alpha,1}(v;s)\Phi_{\alpha,2}(u;s)}{2\pi i(u-v)}.
    \end{align}
\end{theorem}

\begin{remark} \label{rem:equilibrium}
    The equilibrium measure of $S_V = \supp(\psi_V)$ is the unique
    probability measure $\omega_{S_V}$ on $S_V$ that minimizes the logarithmic energy
    \[
        I(\mu) = \iint \log \frac{1}{|x-y|} d\mu(x) d\mu(y)
    \]
    among all probability measures on $S_V$. Since $S_V$ consists of
    a finite union of intervals, and since $0$ is an interior point of one
    of these intervals, $\omega_{S_V}$ has a density $w_{S_V}$ with respect
    to Lebesgue measure, and $w_{S_V}(0) > 0$.
    This number is used in (\ref{definition: s}).
\end{remark}

\begin{remark}
    One can refine the calculations of Section
    \ref{section: proofmainresult} to obtain the following stronger result:
    \begin{equation}
    \frac{1}{c n^{1/3}}
            K_{n,N}\left(\frac{u}{cn^{1/3}}, \frac{v}{cn^{1/3}}\right)
            =
            K^{crit,\alpha}(u,v;s) +\bigO\left(\frac{|u|^\alpha|v|^{\alpha}}{n^{1/3}}\right),
    \end{equation}
    uniformly for $u, v$ in bounded subsets of $\mathbb R\setminus\{0\}$.
\end{remark}

\begin{remark}
    It is not immediate from the expression (\ref{Kcrit}) that
    $K^{crit,\alpha}$ is real. This property follows from the symmetry
    \[
        e^{\frac{1}{2}\pi i \alpha \sgn(u)} \Phi_{\alpha,2}(u;s) =
        \overline{e^{\frac{1}{2} \pi i \alpha \sgn(u)} \Phi_{\alpha,1}(u;s)},
        \qquad \mbox{for $u \in \mathbb R\setminus\{0\}$,}
    \]
    which leads to the ``real formula''
    \[
        K^{crit,\alpha}(u,v;s) =
        - \frac{1}{\pi(u-v)} \Im \left( e^{\frac{1}{2}\pi i \alpha (\sgn(u)-\sgn(v))} \Phi_{\alpha,1}(u;s)
            \overline{\Phi_{\alpha,1}(v;s)} \right),
    \]
    see Remark \ref{remark-reelekern} below.
\end{remark}

\begin{remark}
    For $\alpha = 0$, the theorem is proven in \cite{CK}.
    The proof for the general case follows along similar lines,
    but we need the information about the existence of
    $q_{\alpha}(s)$ for real $s$, as guaranteed by
    Theorem \ref{theorem: polen q}.
\end{remark}

\subsection{Recurrence coefficients for orthogonal polynomials}

In order to prove Theorem \ref{theorem: main result}, we will study the
Riemann-Hilbert problem for orthogonal polynomials with respect to the weight
$|x|^{2\alpha}e^{-NV(x)}$. This analysis leads to asymptotics for the kernel
$K_{n,N}$, but also provides the ingredients to derive asymptotics for the
orthogonal polynomials and for the coefficients in the recurrence relation that
is satisfied by them.

To state these results we introduce measures $\nu_t$
in the following way, see also \cite{CK} and Section 3.2.
Take $\delta_0>0$ sufficiently small and let $\nu_t$
be the minimizer of $I_{V/t}(\nu)$ (see (\ref{defIV}) for the
definition of $I_V$) among all measures
$\nu=\nu^+-\nu^-$, where $\nu^\pm$ are nonnegative measures on
$\mathbb{R}$ such that $\nu(\mathbb{R})=1$ and $\supp (\nu^-)
\subset [-\delta_0, \delta_0]$. We use $\psi_t$ to denote
the density of $\nu_t$.

We restrict ourselves to the one-interval case without
singular points except for $0$.
Then $\supp(\psi_V)=[a,b]$ and $\supp(\psi_t)=[a_t,b_t]$ for $t$ close to $1$,
where $a_t$ and $b_t$ are real analytic functions of $t$.

We write $\pi_{n,N}$ for the monic orthogonal polynomial of degree
$n$ with respect to the weight $|x|^{2\alpha}e^{-NV(x)}$. Those
polynomials satisfy a three-term recurrence relation
\begin{equation}\label{recursie}
    \pi_{n+1,N}=(z-b_{n,N})\pi_{n,N}-a_{n,N}^2\pi_{n-1,N},
\end{equation}
with recurrence coefficients $a_{n,N}$ and $b_{n,N}$. In the large $n$ expansion of $a_{n,N}$
and $b_{n,N}$, we observe oscillations in the $\bigO(n^{-1/3})$-term. The amplitude of the
oscillations is proportional to $q_{\alpha}(s)$,  while in general the frequency of the
oscillations slowly varies with $t=n/N$.
+
\begin{theorem}\label{theorem: recursie}
    Let the conditions of Theorem {\rm \ref{theorem: main result}}
    be satisfied and assume that $\supp(\psi_V)=[a,b]$
    consists of one single interval.
    Consider the
    three-term recurrence relation {\rm (\ref{recursie})} for the monic
    orthogonal polynomials $\pi_{k,N}$ with respect to the weight
    $|x|^{2\alpha}e^{-NV(x)}$. Then as $n, N \to \infty$ such that
    $n/N - 1 = \bigO(n^{-2/3})$, we have
    \begin{equation} \label{recursieanN}
        a_{n,N} =
                \frac{b-a}{4}
                - \frac{q_{\alpha}(s_{t,n})\cos(2\pi n\omega_t +2\alpha\theta)}{2c} n^{-1/3}
                + \bigO(n^{-2/3}),
    \end{equation}
    \begin{equation} \label{recursiebnN}
        b_{n,N} =
                \frac{b+a}{2}
                + \frac{q_{\alpha}(s_{t,n})\sin(2 \pi n\omega_t+(2\alpha+1)\theta)}{c} n^{-1/3}
                + \bigO(n^{-2/3}),
    \end{equation}
    where $t=n/N$, $c$ is given by {\rm (\ref{definition: c})},
    \begin{equation} \label{definition: s recursie}
        s_{t,n} = n^{2/3}\frac{\pi}{c}\psi_t(0),
    \end{equation}
    \begin{equation} \label{definition: theta}
        \theta = \arcsin \frac{b+a}{b-a},
    \end{equation}
    and
    \begin{equation} \label{definition: omegat}
        \omega_t = \int_0^{b_t}\psi_t(x)dx.
    \end{equation}
\end{theorem}

\begin{remark}
    It was shown in \cite{CK} that
    $\left. \frac{d}{dt}\psi_t(0) \right|_{t=1} = w_{S_V}(0)$, which in the
    situation of Theorem \ref{theorem: recursie} implies that (since $S_V = [a,b]$
    and $\psi_t(0)$ is real analytic as a function of $t$ near $t=1$),
    \[
        \psi_t(0) = (t-1) \frac{1}{\pi \sqrt{-ab}} + \bigO((t-1)^2),
        \qquad \mbox{as $t \to 1$.}
    \]
    Then it follows from (\ref{definition: s recursie}) that
    $s_{t,n} = n^{2/3} (t-1) \frac{1}{c \sqrt{-ab}} + \bigO(n^{-2/3})$
    and we could in fact replace $s_{t,n}$ in (\ref{recursieanN})
    and (\ref{recursiebnN}) by
    \[
        s_{t,n}^* = n^{2/3} (t-1) \frac{1}{c \sqrt{-ab}}.
    \]
    We prefer to use $s_{t,n}$ since it appears more naturally from our analysis.
\end{remark}

\begin{remark}
    In \cite{BI2}, Bleher and Its derived (\ref{recursieanN}) in the
    case where $\alpha = 0$ and where $V$ is a critical even quartic
    polynomial. They also computed the $\bigO(n^{-2/3})$-term in
    the large $n$ expansion for $a_{n,N}$. For even $V$ we have that
    $a=-b$, $\theta = 0$, $\omega_t=1/2$ and thus
    $\cos(2\pi n\omega_t + 2 \alpha \theta)=(-1)^n$, so that
    (\ref{recursieanN}) reduces to
    \[
        a_{n,N} =
                \frac{b}{2}-\frac{q_{\alpha}(s_{t,n}) (-1)^n}{2c} n^{-1/3}
                + \bigO(n^{-2/3}),
    \]
    which is in agreement with the result of \cite{BI2}. Also for even $V$ the
    recurrence coefficient $b_{n,N}$ vanishes which is in agreement with
    (\ref{recursiebnN}).
\end{remark}

\begin{remark}
    In \cite{BE} an ansatz was made about the recurrence coefficients
    associated with a general (not necessarily even) critical quartic
    polynomial $V$ in the case $\alpha = 0$. For fixed large $N$, the ansatz agrees
    with (\ref{recursieanN}) and (\ref{recursiebnN}) up to an $N$-dependent
    phase shift in the trigonometric functions.
\end{remark}

\subsection{Outline of the rest of the paper}

In Section 2, we comment on the Riemann-Hilbert problem associated with the
Painlev\'e II equation. We also prove the existence of a solution to this RH
problem for real values of the parameter $s$, and this existence provides the
proof of Theorem \ref{theorem: polen q}. In Section 3, we state the RH problem
for orthogonal polynomials and apply the Deift/Zhou steepest descent method.
Our main focus will be the construction of a local parametrix near the origin.
For this construction, we will use the RH problem from Section 2. In Section 4
and Section 5 finally, we use the results obtained in Section 3 to prove
Theorem \ref{theorem: main result} and Theorem \ref{theorem: recursie}.

\section{RH problem for Painlev\'e II and the proof
of Theorem \ref{theorem: polen q}}
    \label{section: Painleve II}

As before, we assume $\alpha > -1/2$.
\subsection{Statement of RH problem}

\begin{figure}[t]
\begin{center}
    \setlength{\unitlength}{1truemm}
    \begin{picture}(100,48.5)(0,2.5)
        \put(72,27.5){\small $S_1$}
        \put(49,42.5){\small $S_2$}
        \put(27.5,27.5){\small $S_3$}
        \put(49,12.5){\small $S_4$}

        \put(80,47.5){\small $\Gamma_1$}
        \put(17.5,47.5){\small $\Gamma_2$}
        \put(17.5,5){\small $\Gamma_3$}
        \put(80,5){\small $\Gamma_4$}

        \put(50,27.5){\thicklines\circle*{.8}}
        \put(49,30){\small 0}
        \multiput(50,27.5)(3,0){6}{\line(1,0){1.75}}
        \qbezier(60.5,27.5)(61,30.5)(59,32)
        \put(62.25,30){\scriptsize $\pi/6$}

        \put(15,10){\line(2,1){70}}
        \put(15,45){\line(2,-1){70}}
        \put(70,37.5){\thicklines\vector(2,1){.0001}}
        \put(70,17.5){\thicklines\vector(2,-1){.0001}}
        \put(30,37.5){\thicklines\vector(-2,1){.0001}}
        \put(30,17.5){\thicklines\vector(-2,-1){.0001}}
    \end{picture}
    \caption{The contour $\Sigma$ consisting of four straight rays oriented to infinity.}
    \label{figure: RHP Psi}
\end{center}
\end{figure}
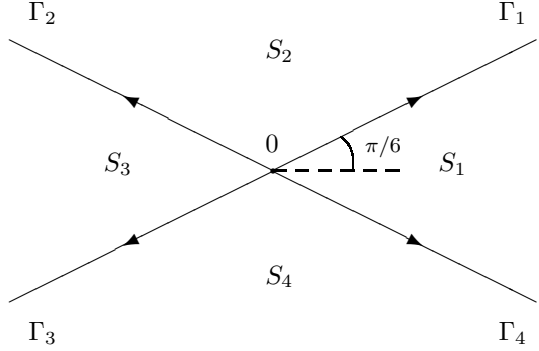

Let $\Sigma = \bigcup_j \Gamma_j$ be the contour consisting
of four straight rays oriented to infinity,
\[
    \Gamma_1:\arg\zeta=\frac{\pi}{6}, \qquad \Gamma_2:\arg\zeta=\frac{5\pi}{6}, \qquad
    \Gamma_3:\arg\zeta=-\frac{5\pi}{6}, \qquad \Gamma_4:\arg\zeta=-\frac{\pi}{6}.
\]
The contour $\Sigma$ divides the complex plane into four regions
$S_1, \ldots, S_4$ as shown in Figure \ref{figure: RHP Psi}.
For $\alpha > -1/2$ and $s \in \mathbb C$, we seek a
$2\times 2$ matrix valued function
$\Psi_{\alpha}(\zeta;s)=\Psi_{\alpha}(\zeta)$
(we suppress notation of $s$ for brevity) satisfying the following.

\subsubsection*{RH problem for $\Psi_{\alpha}$:}

\begin{itemize}
\item[(a)] $\Psi_{\alpha}$ is analytic in $\mathbb C\setminus\Sigma$.
\item[(b)] $\Psi_{\alpha}$ satisfies the following jump
    relations on $\Sigma \setminus \{0\}$,
    \begin{align}
        \label{RHP Psi: b1}
        \Psi_{\alpha,+}(\zeta) &= \Psi_{\alpha,-}(\zeta)
            \begin{pmatrix}
                1 & 0 \\
                e^{-\pi i \alpha} & 1
            \end{pmatrix},
            \qquad \mbox{for $\zeta\in\Gamma_1$,}
        \\[1ex]
        \label{RHP Psi: b2}
        \Psi_{\alpha,+}(\zeta) &= \Psi_{\alpha,-}(\zeta)
            \begin{pmatrix}
                1 & 0 \\
                -e^{\pi i \alpha} & 1
            \end{pmatrix},
        \qquad \mbox{for $\zeta\in\Gamma_2$,}
        \\[1ex]
        \label{RHP Psi: b3}
            \Psi_{\alpha,+}(\zeta) &= \Psi_{\alpha,-}(\zeta)
            \begin{pmatrix}
                1 & e^{-\pi i \alpha} \\
                0 & 1
            \end{pmatrix},
            \qquad \mbox{for $\zeta\in\Gamma_3$,}
        \\[1ex]
        \label{RHP Psi: b4}
            \Psi_{\alpha,+}(\zeta) &= \Psi_{\alpha,-}(\zeta)
            \begin{pmatrix}
                1 & -e^{\pi i \alpha} \\
                0 & 1
            \end{pmatrix},
            \qquad \mbox{for $\zeta\in\Gamma_4$.}
    \end{align}
\item[(c)] $\Psi_{\alpha}$ has the following behavior at infinity,
    \begin{equation}\label{RHP Psi: c}
        \Psi_{\alpha}(\zeta)=(I+\bigO(1/\zeta))e^{-i(\frac{4}{3}\zeta^3+s\zeta)\sigma_3},
        \qquad \mbox{as $\zeta\to\infty$.}
    \end{equation}
    Here $\sigma_3 = \left(\begin{smallmatrix} 1 & 0 \\ 0 & -1 \end{smallmatrix}\right)$
    denotes the third Pauli matrix.
\item[(d)] $\Psi_{\alpha}$ has the following behavior near the origin. If $\alpha<0$,
    \begin{equation}\label{RHP Psi: d1}
        \Psi_{\alpha}(\zeta)=
        \bigO\begin{pmatrix}
            |\zeta|^\alpha & |\zeta|^\alpha \\
            |\zeta|^\alpha & |\zeta|^\alpha
        \end{pmatrix},
        \qquad \mbox{as $\zeta\to 0$,}
    \end{equation}
    and if $\alpha\geq 0$,
    \begin{equation}\label{RHP Psi: d2}
        \Psi_{\alpha}(\zeta)=
        \begin{cases}
            \bigO\begin{pmatrix}
                |\zeta|^{-\alpha} & |\zeta|^{-\alpha} \\
                |\zeta|^{-\alpha} & |\zeta|^{-\alpha}
            \end{pmatrix},
                & \mbox{as $\zeta\to 0,\, \zeta \in S_1\cup S_3$,}
            \\[3ex]
            \bigO\begin{pmatrix}
                |\zeta|^{\alpha} & |\zeta|^{-\alpha} \\
                |\zeta|^{\alpha} & |\zeta|^{-\alpha}
            \end{pmatrix},
                & \mbox{as $\zeta\to 0,\, \zeta \in S_2$,}
            \\[3ex]
            \bigO\begin{pmatrix}
                |\zeta|^{-\alpha} & |\zeta|^{\alpha} \\
                |\zeta|^{-\alpha} & |\zeta|^{\alpha}
            \end{pmatrix},
                & \mbox{as $\zeta\to 0,\, \zeta \in S_4$.}
        \end{cases}
    \end{equation}
\end{itemize}
Note that $\Psi_{\alpha}$ depends on $s$ only through the asymptotic
condition (\ref{RHP Psi: c}).

\begin{remark}
    This RH problem is a generalization of the RH problem for the case
    where $\alpha=0$, used in \cite{BDJ,CK}.
\end{remark}

\begin{remark}
    By standard arguments based on Liouville's theorem, see e.g.~\cite{Deift,KMVV},
    it can be verified that the solution of this RH problem, if it exists, is unique.
    Here it is important that $\alpha > -1/2$.
\end{remark}

In the following we need more information on the behavior of
solutions of the RH problem near $0$.
To this end, we make use of
the following proposition, cf.\ \cite{IK2}.
We use $G_j$ to denote the jump matrix of $\Psi_{\alpha}$ on $\Gamma_j$ as given
by (\ref{RHP Psi: b1})--(\ref{RHP Psi: b4}).

\begin{proposition}\label{proposition: RHP}
    Let $\Psi$ satisfy conditions {\rm (a)}, {\rm (b)}, and {\rm (d)}
    of the RH problem for $\Psi_{\alpha}$.
    \begin{itemize}
        \item[\rm (1)] If $\alpha - \frac{1}{2} \notin\mathbb{N}_0$, then
            there exists an analytic matrix valued function $E$ and constant
            matrices $A_j$ such that
            \begin{equation} \label{definition: Phi}
                \Psi(\zeta) = E(\zeta)
                    \begin{pmatrix}
                        \zeta^{-\alpha} & 0 \\
                        0 & \zeta^{\alpha}
                    \end{pmatrix} A_j,
                    \qquad \mbox{for $\zeta\in S_j$,}
            \end{equation}
            where the branch cut of $\zeta^\alpha$ is chosen along $\Gamma_4$.
            The  matrices $A_j$ satisfy
            \begin{equation} \label{definition: A1: eq1}
                A_{j+1} = A_j G_j,
                    \qquad \mbox{for $j=1,2,3$,}
            \end{equation}
            and
            \begin{equation} \label{definition: A2: eq1}
                A_2=\begin{pmatrix}
                        0 & -p^{-1} \\
                        p & \frac{p}{2\cos (\pi \alpha)}
                    \end{pmatrix},
                    \qquad \mbox{for some $p \in \mathbb C \setminus \{0\}$.}
            \end{equation}
        \item[\rm (2)] If $\alpha- \frac{1}{2} \in \mathbb{N}_0$,
            then there is logarithmic behavior of $\Psi$ at the origin. There exists
            an analytic matrix valued function $E$ and constant matrices $A_j$ such that
            \begin{equation}
                \Psi(\zeta)=E(\zeta)
                    \begin{pmatrix}
                        \zeta^{-\alpha} & 0 \\
                        \frac{1}{\pi} \zeta^\alpha \ln\zeta & \zeta^\alpha
                    \end{pmatrix} A_j,
                    \qquad \mbox{for $\zeta\in S_j$,}
            \end{equation}
            where again the branch cuts of $\zeta^\alpha$
            and $\ln\zeta$ are chosen along $\Gamma_4$.  The matrices $A_j$ satisfy
            \begin{equation} \label{definition: A1: eq2}
                A_{j+1} = A_j G_j,
                    \qquad \mbox{for $j=1,2,3$,}
            \end{equation}
            and for some $p \in \mathbb C$,
            \begin{equation} \label{definition: A2: eq2}
                A_2 =
                \begin{cases}
                    \begin{pmatrix}
                        0 & -1 \\
                        1 & p
                    \end{pmatrix}, & \mbox{if $\alpha-\frac{1}{2}$ is even,}
                \\[3ex]
                    \begin{pmatrix}
                        0 & i \\
                        i & p
                    \end{pmatrix}, & \mbox{if $\alpha -\frac{1}{2}$ is odd.}
                \end{cases}
            \end{equation}
    \end{itemize}
\end{proposition}

\begin{proof}
    (1)
    Define $E$ by equation (\ref{definition: Phi}) with matrices $A_j$
    satisfying (\ref{definition: A1: eq1}) and (\ref{definition: A2: eq1}).
    Then $E$ is analytic across
    $\Gamma_1$, $\Gamma_2$, and $\Gamma_3$ because of (\ref{definition: A1: eq1}).
    For $\zeta \in \Gamma_4$ there is a jump
    \begin{equation} \label{eq:jumpE}
        E_+(\zeta) = E_-(\zeta)
            \begin{pmatrix}
                \zeta^{-\alpha} & 0 \\
                0 & \zeta^{\alpha}
            \end{pmatrix}_-
            A_4 G_4 A_1^{-1}
            \begin{pmatrix}
                \zeta^{\alpha} & 0 \\
                0 & \zeta^{-\alpha}
            \end{pmatrix}_+.
        \end{equation}
    Using $\zeta^{\alpha}_- = e^{2\pi i\alpha}\zeta^{\alpha}_+$ and the explicit
    expressions for the matrices $G_j$ and $A_j$, we get from (\ref{eq:jumpE})
    that $E$ is analytic across $\Gamma_4$ as well.

    What remains to be shown is that the possible isolated singularity
    of $E$ at the origin is removable.
    If $\alpha <0$ it follows from (\ref{RHP Psi: d1}) and (\ref{definition: Phi}) that
    \[
        E(\zeta)=\bigO
        \begin{pmatrix}
            |\zeta|^{2\alpha} & 1 \\
            |\zeta|^{2\alpha} & 1
        \end{pmatrix}, \qquad \mbox{as $\zeta\to 0$,}
    \]
    so that (since $2\alpha>-1$) the isolated singularity at the origin is
    indeed removable. If $\alpha>0$ we have in sector $S_2$ by
    (\ref{RHP Psi: d2}), (\ref{definition: Phi}), and
    (\ref{definition: A2: eq1}) that
    \[
        E(\zeta) =\Psi(\zeta) A_2^{-1}
        \begin{pmatrix}
            \zeta^{\alpha} & 0 \\
            0 & \zeta^{-\alpha}
        \end{pmatrix}
        = \bigO\begin{pmatrix}
            |\zeta|^\alpha & |\zeta|^{-\alpha} \\
            |\zeta|^\alpha & |\zeta|^{-\alpha}
        \end{pmatrix}
        \begin{pmatrix}
            * & * \\
            * & 0
        \end{pmatrix}
        \begin{pmatrix}
            \zeta^{\alpha} & 0 \\
            0 & \zeta^{-\alpha}
        \end{pmatrix}
        = \bigO\begin{pmatrix}
            1 & 1 \\
            1 & 1
        \end{pmatrix},
    \]
    as $\zeta\to 0,\, \zeta\in S_2$, where $*$ denotes an unimportant constant.
    Hence the singularity at the origin is not a pole. Moreover, from
    (\ref{RHP Psi: d2}) and (\ref{definition: Phi}) it is also
    easy to check that $E$ does not have an essential singularity at
    the origin either. Therefore the singularity is  removable for
    the case $\alpha>0$ as well, and the proof of part (1) is complete.
    \medskip

    (2) The proof of part (2) is similar.
\end{proof}

\begin{remark}
    The matrix $A_2$ in Proposition \ref{proposition: RHP} is called the
    connection matrix, cf.\ \cite{FlaschkaNewell,FokasZhou}. In all cases
    we have $\det A_2 = 1$ and the $(1,1)$-entry of $A_2$ is zero.
\end{remark}

\subsection{Solvability of the RH Problem for $\Psi_{\alpha}$}

We are going to prove that the RH problem for $\Psi_{\alpha}$ is solvable for
every $s \in \mathbb R$. We do that, as in \cite{DKMVZ2,FokasZhou,Zhou},
by showing that every solution of the homogeneous RH problem is
identically zero. Such a result is known as a vanishing lemma
\cite{FokasMuganZhou,FokasZhou}.

We briefly indicate why the vanishing lemma is enough to establish the
solvability of the RH problem for $\Psi_{\alpha}$. The RH problem is equivalent
to a singular integral equation on the contour $\Sigma$. The singular
integral equation can be stated in operator theoretic terms, and the
operator is a Fredholm operator of zero index. The vanishing lemma yields
that the kernel is trivial, and so the operator is onto which implies
that the singular integral equation is solvable, and therefore the
RH problem is solvable.
For more details and other examples of this procedure
see \cite{DKMVZ2,FokasZhou,Zhou} and \cite[Appendix A]{KMM}.

\begin{proposition} (\textbf{vanishing lemma}) \label{prop:vanishinglemma}
    Let $\alpha > -1/2$ and $s \in \mathbb R$.
    Suppose that $\widehat\Psi$ satisfies conditions {\rm (a)}, {\rm (b)}, and
    {\rm (d)} of the RH problem for $\Psi_{\alpha}$ with the following
    asymptotic condition (instead of condition {\rm (c)})
    \begin{equation} \label{new asymptotic condition}
        \widehat\Psi(\zeta)e^{i(\frac{4}{3}\zeta^3+s\zeta)\sigma_3}=\bigO(1/\zeta),
        \qquad \mbox{as $\zeta\to\infty$.}
    \end{equation}
    Then $\widehat{\Psi} \equiv 0$.
\end{proposition}

\begin{proof}
As before, we use $G_j$ to denote
the jump matrix of $\Gamma_j$, given by (\ref{RHP Psi: b1})--(\ref{RHP Psi: b4}).
Introduce an auxiliary matrix valued function $H$ with jumps only
on $\mathbb R$, as follows.
\begin{equation}\label{definition: H}
    H(\zeta)=
    \begin{cases}
        \widehat\Psi(\zeta)e^{i(\frac{4}{3}\zeta^3+s\zeta)\sigma_3},
            & \mbox{for $\zeta\in S_2\cup S_4$,}
        \\[2ex]
        \widehat\Psi(\zeta)G_1 e^{i(\frac{4}{3}\zeta^3+s\zeta)\sigma_3},
            & \mbox{for $\zeta\in S_1\cap\mathbb C_+$,}
        \\[2ex]
        \widehat\Psi(\zeta)G_2^{-1} e^{i(\frac{4}{3}\zeta^3+s\zeta)\sigma_3},
            & \mbox{for $\zeta\in S_3\cap\mathbb C_+$,}
        \\[2ex]
        \widehat\Psi(\zeta)G_3 e^{i(\frac{4}{3}\zeta^3+s\zeta)\sigma_3},
            & \mbox{for $\zeta\in S_3\cap\mathbb C_-$,}
        \\[2ex]
        \widehat\Psi(\zeta)G_4^{-1} e^{i(\frac{4}{3}\zeta^3+s\zeta)\sigma_3},
            & \mbox{for $\zeta\in S_1\cap\mathbb C_-$.}
    \end{cases}
\end{equation}
Then $H$ satisfies the following RH problem.

\subsubsection*{RH problem for $H$:}

\begin{itemize}
    \item[(a)] $H:\mathbb C\setminus\mathbb R\to\mathbb C^{2\times 2}$ is
        analytic and satisfies the following jump relations on
        $\mathbb R\setminus\{0\}$,
        \begin{align}\label{RHP_H: b1}
            H_+(\zeta)&=H_-(\zeta)e^{-i(\frac{4}{3}\zeta^3+s\zeta)\sigma_3}
                \begin{pmatrix}
                    0 & -e^{-\pi i\alpha} \\
                    e^{\pi i\alpha} & 1
                \end{pmatrix} e^{i(\frac{4}{3}\zeta^3+s\zeta)\sigma_3},\quad \mbox{for $\zeta\in(-\infty,0)$,} \\[1ex]
            \label{RHP_H: b2}
            H_+(\zeta)&=H_-(\zeta)e^{-i(\frac{4}{3}\zeta^3+s\zeta)\sigma_3}
                \begin{pmatrix}
                    0 & -e^{\pi i\alpha} \\
                    e^{-\pi i\alpha} & 1
                \end{pmatrix} e^{i(\frac{4}{3}\zeta^3+s\zeta)\sigma_3},\quad \mbox{for $\zeta\in(0,\infty)$.}
        \end{align}
    \item[(b)] $H(\zeta)=\bigO(1/\zeta)$, \qquad as $\zeta\to\infty$.
    \item[(c)] $H$ has the following behavior near the origin: If
        $\alpha<0$,
        \begin{equation} \label{eq:Hnear0-1}
            H(\zeta)=\bigO\begin{pmatrix}
                |\zeta|^\alpha & |\zeta|^\alpha \\
                |\zeta|^\alpha & |\zeta|^\alpha
            \end{pmatrix},\qquad \mbox{as $\zeta\to 0$,}
        \end{equation}
        and if $\alpha>0$,
        \begin{equation} \label{eq:Hnear0-2}
            H(\zeta)=\begin{cases}
                \bigO\begin{pmatrix}
                    |\zeta|^\alpha & |\zeta|^{-\alpha} \\
                    |\zeta|^\alpha & |\zeta|^{-\alpha}
                \end{pmatrix}, & \mbox{as $\zeta\to 0, \Im
                \zeta>0$,} \\[3ex]
                \bigO\begin{pmatrix}
                    |\zeta|^{-\alpha} & |\zeta|^\alpha \\
                    |\zeta|^{-\alpha} & |\zeta|^\alpha
                \end{pmatrix}, & \mbox{as $\zeta\to 0, \Im
                \zeta<0$.}
            \end{cases}
        \end{equation}
\end{itemize}

The jumps in (a) follow from straightforward calculation.
The vanishing behavior (b) of $H$ at infinity (in all sectors) follows from
the triangular shape of the jump matrices $G_j$, see (\ref{RHP Psi: b1})--(\ref{RHP Psi: b4}).
For example, for $\zeta\in S_1\cap\mathbb C_+$ we have $\Re i(\frac{4}{3}\zeta^3+s\zeta)<0$ so
that by (\ref{new asymptotic condition}) and (\ref{definition: H})
\[
    H(\zeta)
        = \bigO(1/\zeta)
        \begin{pmatrix}
            1 & 0 \\
            e^{-\pi i\alpha}e^{2i(\frac{4}{3}\zeta^3+s\zeta)} & 1
        \end{pmatrix}
        = \bigO(1/\zeta),\qquad \mbox{as $\zeta\to\infty$.}
\]
The behavior near the origin in (c) follows from Proposition \ref{proposition:
RHP}. This is immediate for (\ref{eq:Hnear0-1}), while for $\alpha > 0$,
$\alpha - \frac{1}{2} \not\in \mathbb N_0$, we have by  (\ref{definition:
Phi}), (\ref{definition: A1: eq1}), (\ref{definition: A2: eq1}), and
(\ref{definition: H}),
\[
    H(\zeta)e^{-i(\frac{4}{3}\zeta^3+s\zeta)\sigma_3} =
    \begin{cases}
        E(\zeta) \zeta^{-\alpha\sigma_3} A_2
            = E(\zeta) \zeta^{-\alpha\sigma_3}
            \begin{pmatrix}
                0 & * \\
                * & *
            \end{pmatrix},
            & \mbox{if $\Im \zeta  > 0$,}
        \\[3ex]
        E(\zeta) \zeta^{-\alpha\sigma_3} A_4
            = E(\zeta) \zeta^{-\alpha\sigma_3}
            \begin{pmatrix}
                * & 0 \\
                * & *
            \end{pmatrix},
            & \mbox{if $\Im \zeta < 0$,}
    \end{cases}
\]
which yields (\ref{eq:Hnear0-2}) in case $\alpha - \frac{1}{2}
\not\in \mathbb N_0$, since $E$ is analytic.
Using (\ref{definition: A2: eq2}) instead of
(\ref{definition: A2: eq1}), we will see that the same
argument works if $\alpha - \frac{1}{2} \in \mathbb N_0$.
\medskip

Next we define, cf.\ \cite{DKMVZ2,FokasZhou,Zhou},
\begin{equation} \label{eq:defM}
    M(\zeta)=H(\zeta)H(\bar\zeta)^*,\qquad \mbox{for
    $\zeta\in\mathbb C\setminus\mathbb R$,}
\end{equation}
where $H^*$ denotes the Hermitian conjugate of $H$. From condition (c)
of the RH problem for $H$ it follows that $M$ has the following
behavior near the origin
\[
    M(\zeta)=
    \begin{cases}
        \bigO\begin{pmatrix}
            |\zeta|^{2\alpha} & |\zeta|^{2\alpha} \\
            |\zeta|^{2\alpha} & |\zeta|^{2\alpha}
        \end{pmatrix}, & \mbox{as $\zeta\to 0$, in case
        $\alpha<0$,} \\[3ex]
        \bigO\begin{pmatrix}
            1 & 1 \\
            1 & 1
        \end{pmatrix},& \mbox{as $\zeta\to 0$, in case
        $\alpha>0$.}
    \end{cases}
\]
Since $\alpha>-1/2$, it follows that each entry of $M$ has an integrable singularity at the
origin. Because $M(\zeta)=\bigO(1/\zeta^2)$ as $\zeta\to\infty$, and $M$ is analytic in the
upper half plane,  it then follows by Cauchy's theorem that $\int_{\mathbb R}
M_+(\zeta)d\zeta=0$, and hence by (\ref{eq:defM})
\[
    \int_{\mathbb R} H_+(\zeta)H_-(\zeta)^*d\zeta=0.
\]
Adding this equation to its Hermitian conjugate, we find
\begin{equation}\label{int H-}
    \int_{\mathbb R} \left[ H_+(\zeta) H_-(\zeta)^* + H_-(\zeta) H_+(\zeta)^*\right] d\zeta = 0.
\end{equation}
Using (\ref{RHP_H: b1}), (\ref{RHP_H: b2}) and the fact that
$(e^{i(\frac{4}{3}\zeta^3+s\zeta)\sigma_3})^*=e^{-i(\frac{4}{3}\zeta^3+s\zeta)\sigma_3}$
for $\zeta,s\in\mathbb R$, (here we use the fact that $s$ is real!),
we obtain from (\ref{int H-}),
\[
    0   = \int_{\mathbb R} H_-(\zeta) \begin{pmatrix}
        0 & 0 \\
        0 & 2
    \end{pmatrix}H_-(\zeta)^*d\zeta
        = 2 \int_{\mathbb R}
            \left[ \left|(H_-)_{12}(\zeta)\right|^2 +
                   \left|(H_-)_{22}(\zeta)\right|^2 \right]
                   d\zeta.
\]
This implies that the second column of $H_-$ is identically zero.
The jump relations (\ref{RHP_H: b1}) and (\ref{RHP_H: b2}) of $H$
then imply that the first column of $H_+$ is identically zero as
well.

\medskip

To show that the second column of $H_+$ and the first
column of $H_-$ are also identically zero, we use an idea of Deift
et al.\ \cite[Proof of Theorem 5.3, Step 3]{DKMVZ2}.
Since the second column of $H_-$ is identically zero, the jump
relations (\ref{RHP_H: b1}) and (\ref{RHP_H: b2}) for $H$ yield
for $j=1,2$,
\[
    \left( H_{j2} \right)_+(\zeta) = - e^{ \sgn(\zeta) \pi i \alpha}
    e^{-2i(\frac{4}{3}\zeta^3+s\zeta)} \left(H_{j1}\right)_-(\zeta),
    \qquad \mbox{for $\zeta \in \mathbb R \setminus \{0\}$.}
\]
Thus if we define for $j=1,2$,
\begin{equation}\label{definition: h}
    h_j(\zeta)=
        \begin{cases}
            H_{j2}(\zeta), & \mbox{ if $\Im \zeta >0$}, \\
            H_{j1}(\zeta),  & \mbox{ if $\Im \zeta <0$},
        \end{cases}
\end{equation}
then both $h_1$ and $h_2$ satisfy the following RH problem
for a scalar function $h$.

\subsubsection*{RH problem for $h$:}

\begin{itemize}
    \item[(a)] $h$ is analytic on $\mathbb C\setminus\mathbb R$ and
        satisfies the following jump relation
        \[ h_+(\zeta) = -e^{ \sgn(\zeta) \pi
            i\alpha}e^{-2i(\frac{4}{3}\zeta^3+s\zeta)}  h_-(\zeta),
            \qquad \mbox{for $\zeta\in \mathbb R \setminus \{0\}$},
        \]
    \item[(b)] $h(\zeta)=\bigO(1/\zeta)$ as $\zeta\to\infty$.
    \item[(c)] $h(\zeta)=\begin{cases}
                \bigO(|\zeta|^\alpha),
                    & \mbox{as $\zeta\to 0$, in case $\alpha<0$}, \\
                \bigO(|\zeta|^{-\alpha}),
                    & \mbox{as $\zeta\to 0$, in case $\alpha>0$}.
                \end{cases}$
\end{itemize}
Take $\zeta_0$ with $\Im \zeta_0 < -1$ and define
\begin{equation}
    \widehat h(\zeta) =
            \begin{cases}
                \frac{\zeta^{\alpha}}{(\zeta-\zeta_0)^{\alpha}} h(\zeta),
                    & \mbox{if $\Im \zeta>0$,}
                \\[2ex]
                \frac{\zeta^{\alpha}}{(\zeta-\zeta_0)^{\alpha}}
                \left(-e^{\pi i\alpha}e^{-2i(\frac{4}{3}\zeta^3+s\zeta)}\right) h(\zeta),
                    & \mbox{if $-1<\Im \zeta<0$,}
            \end{cases}
\end{equation}
where we use principal branches of the powers, so that
$\zeta^\alpha$ is defined with a branch cut along the negative
real axis. Then it is easy to check that $\widehat h$ is analytic
in $\Im \zeta>-1$, continuous and uniformly bounded in
$\Im\zeta\geq -1$, and
\[
    \widehat h(\zeta)=\bigO(e^{-3|\Re \zeta|^2}),
        \qquad \mbox{as $\zeta\to\infty$ on the horizontal line $\Im \zeta=-1$.}
\]
By Carlson's theorem, see e.g.\ \cite{ReedSimon}, this implies that $\widehat
h\equiv 0$, so that $h\equiv 0$, as well. This in turn implies that $h_1 \equiv
0$ and $h_2 \equiv 0$, so that $H \equiv 0$. Then also $\widehat{\Psi} \equiv
0$ and the proposition is proven.
\end{proof}

As noted before, Proposition \ref{prop:vanishinglemma}
has the following consequence.

\begin{corollary} \label{solvable}
    The RH problem for $\Psi_{\alpha}$, see Section {\rm 2.1},
    has a unique solution for every  $s \in \mathbb R$ and $\alpha > -1/2$.
\end{corollary}

\subsection{Proof of Theorem \ref{theorem: polen q}}

Theorem \ref{theorem: polen q} follows from the
connection of the RH problem for $\Psi_{\alpha}$ of Section 2.1 with the RH problem
associated with the general Painlev\'e II equation (\ref{PIIvgl})
as first described by Flaschka and Newell \cite[Section 3D]{FlaschkaNewell}.

\begin{varproof} \textbf{of Theorem \ref{theorem: polen q}.}
Consider the matrix differential equation
\begin{equation} \label{diffPsiA}
     \frac{\partial \Psi}{\partial \zeta} = A \Psi,
     \end{equation}
where $A$ is as in (\ref{AenB}) and $s$, $q$, and $r$ are constants. For every $k=0,1,\ldots,
5$, there is a unique solution $\Psi_k$ of (\ref{diffPsiA}) such that $\Psi_k(\zeta) = (I +
\bigO(1/\zeta))
    e^{-i(\frac{4}{3} \zeta^3 + s\zeta)\sigma_3}$
as $\zeta \to \infty$ in the sector $ (2k-1) \frac{\pi}{6} <
    \arg \zeta < (2k+1) \frac{\pi}{6}$.
The function
\begin{equation} \label{defpiecewisePsi}
    \Psi(\zeta) = \Psi_k(\zeta),
        \qquad \mbox{for $(2k-1) \frac{\pi}{6} < \arg \zeta < (2k+1)
        \frac{\pi}{6}$,}
    \end{equation}
is then defined on $\mathbb C \setminus (\Sigma \cup i\mathbb R)$
and satisfies the following conditions.
\begin{itemize}
\item[(a)] $\Psi$ is analytic in
    $\mathbb C\setminus(\Sigma\cup i\mathbb R)$.
\item[(b)] There exist constants $s_1, s_2, s_3 \in \mathbb C$
(Stokes multipliers) satisfying
\begin{equation} \label{cycliccond}
    s_1+s_2+s_3+s_1s_2s_3=-2i\sin \pi\alpha
    \end{equation}
 such that the following jump conditions hold, where all
 rays are oriented to infinity,
    \[ \Psi_+=
        \begin{cases}
            \Psi_-
            \begin{pmatrix}
                1 & 0 \\
                s_1 & 1
            \end{pmatrix},
            & \mbox{on $\Gamma_1$,}
            \\[3ex]
            \Psi_-
            \begin{pmatrix}
                1 & s_2 \\
                0 & 1
            \end{pmatrix},
            & \mbox{on $i\mathbb R^+$,}
            \\[3ex]
            \Psi_-
            \begin{pmatrix}
                1 & 0 \\
                s_3 & 1
            \end{pmatrix},
            & \mbox{on $\Gamma_2$,}
        \end{cases}
        \qquad \qquad
        \Psi_+ =
            \begin{cases}
            \Psi_-
            \begin{pmatrix}
                1 & s_1 \\
                0 & 1
            \end{pmatrix},
            & \mbox{on $\Gamma_3$,}
            \\[3ex]
            \Psi_-
            \begin{pmatrix}
                1 & 0 \\
                s_2 & 1
            \end{pmatrix},
            & \mbox{on $i\mathbb R^-$,}
            \\[3ex]
            \Psi_-
            \begin{pmatrix}
                1 & s_3 \\
                0 & 1
            \end{pmatrix},
            & \mbox{on $\Gamma_4$.}
        \end{cases}
    \]
    \item[(c)]
    $\Psi(\zeta)=(I+\bigO(1/\zeta))e^{-i(\frac{4}{3}\zeta^3+s\zeta)\sigma_3}$,\qquad
        as $\zeta\to\infty$.
\end{itemize}

The Stokes multipliers $s_1,s_2, s_3$ depend on $s$, $q$ and $r$.
However, if $q = q(s)$ satisfies the second Painlev\'e equation
$q'' = sq + 2q^3 - \alpha$, and if $r = q'(s)$, then the Stokes multipliers
are constant. In this way there is a one-to-one correspondence between
solutions of the Painlev\'e II equation and Stokes multipliers
$s_1, s_2, s_3$ satisfying (\ref{cycliccond}). This also means that there
exists a solution of the above RH problem which is built out of solutions
of (\ref{diffPsiA}) if and only if $s$ is not a pole of the
Painlev\'e II function that corresponds to the Stokes multipliers
$s_1, s_2, s_3$. The Painlev\'e II function itself may then be recovered from
the RH problem by the formula \cite{FlaschkaNewell}
\[
        q(s)=\lim_{\zeta\to\infty}
            2i\zeta\Psi_{12}(\zeta)e^{-i(\frac{4}{3}\zeta^3+s\zeta)},
    \]
with $\Psi_{12}$ the $(1,2)$-entry of $\Psi$. In particular,
condition (c) of the RH problem can be strengthened to
 \begin{equation}
    \label{Psiinfty}\Psi(\zeta)=\left(I+\frac{1}{2i\zeta}
    \begin{pmatrix}u(s)&q(s)\\-q(s)&-u(s)\end{pmatrix}+\bigO(1/\zeta^2)\right)
    e^{-i(\frac{4}{3}\zeta^3+s\zeta)\sigma_3},\qquad\mbox{as $\zeta\to\infty$,}
    \end{equation}
        where $u = (q')^2 - sq^2 - q^4 + 2 \alpha q$.

The RH problem for $\Psi_{\alpha}$ in Section 2.1 corresponds to
\begin{equation} \label{specialStokes}
    s_1 = e^{-\pi i\alpha}, \qquad s_2 = 0, \qquad s_3 = - e^{\pi i \alpha}.
    \end{equation}
These Stokes multipliers are very special in two respects \cite{IK1,Kap}.
First, since $s_2 = 0$, the corresponding solution of
the Painlev\'e II equation decays as $s \to +\infty$, i.e.,
\begin{equation} \label{decaying}
    q(s) \sim \frac{\alpha}{s}, \qquad \mbox{as $s \to +\infty$.}
    \end{equation}
Secondly, since  $s_1s_3 = -1$ the Painlev\'e II solution increases
as $s \to -\infty$, i.e.,
\begin{equation} \label{increasing}
    q(s) \sim \pm \sqrt{-\frac{s}{2}}\, , \qquad \mbox{as $s \to -\infty$.}
    \end{equation}
where the choice $s_1 = e^{-\pi i\alpha}$, $s_3 = -e^{\pi i \alpha}$
corresponds to the $+$sign, while the interchange of $s_1$ and $s_3$
corresponds to the $-$ sign in (\ref{increasing}).
Thus the special choice (\ref{specialStokes}) corresponds to $q_{\alpha}$,
the Hastings-McLeod solution of the general Painlev\'e II equation,
see (\ref{asymptotics q +infty}) and (\ref{asymptotics q -infty}).

Then as a consequence of the fact that the RH problem for $\Psi_{\alpha}$
stated in Section 2.1 is solvable for every real $s$ by Corollary \ref{solvable},
we conclude that $q_{\alpha}$ has no poles on the real line, which
proves Theorem \ref{theorem: polen q}.
\end{varproof}

\begin{remark}
Its and Kitaev \cite{IK1} use a slightly modified, but equivalent, version
of the RH problem for $\Psi_{\alpha}$. The solutions are connected by the transformation
\begin{equation} \label{IKversion}
\Psi_{\alpha} \leftrightarrow e^{\frac{\pi i}{4} \sigma_3}\Psi_{\alpha}
e^{-\frac{\pi i}{4} \sigma_3},
\end{equation}
which results in a transformaton of the Stokes
multipliers $s_j\leftrightarrow (-1)^j i s_j$.
\end{remark}

For later use, we record the following corollary.
\begin{corollary}
For every fixed $s_0\in\mathbb R$, there exists an open
neighborhood $U$ of $s_0$ such that the RH problem for $\Psi_{\alpha}$ is
solvable for every $s\in U$.
\end{corollary}
\begin{proof}
Since $q_\alpha$ is meromorphic in  $\mathbb C$, there is an open
neighborhood of $s_0$ without poles.  This implies
\cite{FlaschkaNewell} that the RH problem for $\Psi_{\alpha}$ is
solvable for every $s$ in that open neighborhood of $s_0$, as
well.
\end{proof}

\begin{remark}\label{remark: analyticitypsi}
The function $\Psi_{\alpha}(\zeta;s)$ is analytic as a function of
both $\zeta \in \mathbb C \setminus \Sigma$ and $s\in\mathbb
C\setminus\mathcal{P}_{\alpha}$, where $\mathcal{P}_{\alpha}$
denotes the set of poles of $q_\alpha$, see \cite{FlaschkaNewell}.
As a consequence, one can check that (\ref{RHP Psi: c}),
(\ref{RHP Psi: d1}) and (\ref{RHP Psi: d2}) hold uniformly for
$s$ in compact subsets of $\mathbb C\setminus\mathcal{P}_{\alpha}$.
\end{remark}

\begin{remark}
The functions $\Phi_{\alpha,1}$ and $\Phi_{\alpha,2}$
defined by (\ref{systeemdiffvgln}) and (\ref{asymptotiekphi})
are connected with $\Psi_{\alpha}$ as follows.
Define
\begin{equation}\label{definition: Phi_alpha}
    \Phi_{\alpha}(\zeta;s)=
    \begin{cases}
        \Psi_{\alpha}(\zeta;s)
            \begin{pmatrix}
                1 & 0 \\
                e^{-\pi i \alpha} & 1
            \end{pmatrix}, & \mbox{for $\zeta\in S_1$,} \\[3ex]
        \Psi_{\alpha}(\zeta;s), & \mbox{for $\zeta\in S_2$,} \\[2ex]
        \Psi_{\alpha}(\zeta;s)
            \begin{pmatrix}
                1 & 0 \\
                e^{\pi i \alpha} & 1
            \end{pmatrix}, & \mbox{for $\zeta\in S_3$,} \\[3ex]
        \Psi_{\alpha}(\zeta;s)
            \begin{pmatrix}
                1 & -e^{\pi i \alpha} \\
                0 & 1
            \end{pmatrix}
            \begin{pmatrix}
                1 & 0 \\
                e^{-\pi i \alpha} & 1
            \end{pmatrix}, & \mbox{for $\zeta \in S_4$, $\Re \zeta > 0$,} \\[3ex]
        \Psi_{\alpha}(\zeta;s)
            \begin{pmatrix}
                1 & -e^{-\pi i \alpha} \\
                0 & 1
            \end{pmatrix}
            \begin{pmatrix}
                1 & 0 \\
                e^{\pi i \alpha} & 1
            \end{pmatrix}, & \mbox{for $\zeta \in S_4$, $\Re \zeta < 0$.}
    \end{cases}
\end{equation}
Then it follows from the RH problem for $\Psi_{\alpha}$ that
$\Phi_{\alpha}$ is analytic on $\mathbb C \setminus (-i\infty,0]$.
Morever, we also see from (\ref{systeemdiffvgln}) and (\ref{asymptotiekphi})
that
\begin{equation} \label{firstcolumn}
    \Phi_{\alpha} = \begin{pmatrix} \Phi_{\alpha,1} & * \\ \Phi_{\alpha,2}& *
    \end{pmatrix},
    \end{equation}
where $*$ denotes an unspecified unimportant entry. It also follows that
$\Phi_{\alpha,1}$ and $\Phi_{\alpha,2}$ have analytic continuations to
$\mathbb C \setminus (-i\infty,0]$.
\end{remark}

\begin{remark}\label{remark-reelekern}
We show that the kern $K^{crit,\alpha}(u,v;s)$ is real. This will
follow from the identity
\begin{equation} \label{symmetries}
    e^{\frac{1}{2}\pi i \alpha \sgn(u)} \Phi_{\alpha,2}(u;s) =
    \overline{e^{\frac{1}{2} \pi i \alpha \sgn(u)} \Phi_{\alpha,1}(u;s)},
    \qquad \mbox{for } u \in \mathbb R\setminus\{0\} \mbox{ and } s \in \mathbb R,
\end{equation}
since obviously (\ref{symmetries}) implies that
\[ K^{crit,\alpha}(u,v;s) =
    - \frac{1}{\pi(u-v)} \Im \left( e^{\frac{1}{2}\pi i \alpha (\sgn(u)-\sgn(v))} \Phi_{\alpha,1}(u;s)
        \overline{\Phi_{\alpha,1}(v;s)} \right).
\]

The identity (\ref{symmetries}) will follow from the RH problem.
It is easy to check that $\sigma_1\overline{\Psi_{\alpha}(\overline{\zeta};s)}
\sigma_1$, with $\sigma_1= \left(\begin{smallmatrix}
    0 & 1\\
    1 & 0
\end{smallmatrix}\right)$,
also satisfies the RH conditions for $\Psi_{\alpha}$. Because of the uniqueness of
the solution of the RH problem, this implies
\begin{equation} \label{symmetries2}
    \Psi_{\alpha}(\zeta;s) =
        \sigma_1\overline{\Psi_{\alpha}(\overline{\zeta}; s)}\sigma_1.
        \end{equation}
For $\zeta \in S_4$,  the equality of the $(2,1)$ entries
of (\ref{symmetries2}) yields by (\ref{definition: Phi_alpha})
and (\ref{firstcolumn})
\begin{equation} \label{symmetries3}
    e^{\pi i \alpha} \Phi_{\alpha,2}(\zeta; s) = \overline{\Phi_{\alpha,1}(\zeta;s)},
    \qquad \mbox{for } \zeta \in S_4, \ \Re \zeta > 0,
\end{equation}
and
\begin{equation} \label{symmetries4}
    e^{-\pi i \alpha} \Phi_{\alpha,2}(\zeta;s) = \overline{\Phi_{\alpha,1}(\zeta;s)},
    \qquad  \mbox{for } \zeta \in S_4, \ \Re \zeta < 0.
    \end{equation}
Since both sides of (\ref{symmetries3}) are analytic in the right half-plane
we find the identity (\ref{symmetries}) for $u > 0$, and similarly since both
sides of (\ref{symmetries4}) are analytic in the left half-plane, we obtain
(\ref{symmetries}) for $u < 0$.
\end{remark}

\section{Steepest descent analysis of RH problem}
    \label{section: asymptotic analysis}

In this section we write the kernel $K_{n,N}$ in terms of the solution $Y$
of the RH problem for orthogonal polynomials (due to Fokas, Its
and Kitaev \cite{FokasItsKitaev}) and apply the Deift/Zhou
steepest descent method \cite{DeiftZhou} to the RH problem for $Y$
to get the asymptotics for $Y$. These asymptotics will
be used in the next sections to prove
Theorems \ref{theorem: main result} and \ref{theorem: recursie}.

We will restrict ourselves to the one-interval case, which means
that $\psi_V$ is supported on one interval, although the RH
analysis can be done in general. We comment below
in Remark \ref{multiinterval} (see the end of this section)
on the modifications that have to be
made in the multi-interval case.

As in Theorems \ref{theorem: main result} and
\ref{theorem: recursie} we also assume that
besides $0$ there are no other singular points.

\subsection{RH problem for orthogonal polynomials}

The starting point is the RH problem that characterizes the orthogonal
polynomials associated with the weight $|x|^{2\alpha} e^{-NV(x)}$. The $2\times
2$ matrix-valued function $Y=Y_{n,N}$ satisfies the following conditions.
\subsubsection*{RH problem for $Y$}
\begin{itemize}
    \item[(a)] $Y:\mathbb C \setminus \mathbb R \to \mathbb C^{2\times 2}$ is analytic.
    \item[(b)] $Y_+(x)=Y_-(x)
                \begin{pmatrix}
                    1 & |x|^{2\alpha}e^{-NV(x)} \\
                    0 & 1
                \end{pmatrix},$
                \qquad  for $x \in\mathbb R$.
    \item[(c)] $Y(z)=\left(I+\bigO(1/z)\right)
                \begin{pmatrix}
                    z^n & 0 \\
                    0 & z^{-n}
                \end{pmatrix}$,
                \qquad  as $z\to \infty $.
    \item[(d)] $Y$ has the following behavior near the origin,
                \begin{equation}\label{RHP Y: d}
                    Y(z)=\begin{cases}
                    \bigO\begin{pmatrix}
                        1 & |z|^{2\alpha} \\
                        1 & |z|^{2\alpha}
                    \end{pmatrix}, &\mbox{as $z\to 0$, if
                    $\alpha<0$,}\\[3ex]
                    \bigO\begin{pmatrix}
                        1 & 1 \\
                        1 & 1
                    \end{pmatrix}, &\mbox{as $z\to 0$, if $\alpha\geq 0$.}
                    \end{cases}
                \end{equation}
\end{itemize}
Here we have oriented the real axis from the left to the right and
$Y_+(x)$ ($Y_-(x)$) in part (b) denotes the limit as we approach $x \in
\mathbb R$ from the upper (lower) half-plane. This RH problem
possesses a unique solution given by \cite{FokasItsKitaev}
(see \cite{KMVV,KV2} for the condition (d)),
\begin{equation}\label{RHP Y: solution}
    Y(z)=
    \begin{pmatrix}
        {\ds\frac{1}{\kappa_{n,N}} p_{n,N}(z)} &
        {\ds\frac{1}{2\pi i\kappa_{n,N}} \int_{\mathbb R} \frac{p_{n,N}(y)|y|^{2\alpha}e^{- N
        V(y)}}{y-z}dy}
        \\[2ex]
        {\ds -2\pi i \kappa_{n-1,N} p_{n-1,N}(z)} &
        {\ds -\kappa_{n-1,N} \int_{\mathbb R}\frac{p_{n-1,N}(y) |y|^{2\alpha}e^{- N
        V(y)}}{y-z}dy}
\end{pmatrix},
\end{equation}
for $z\in\mathbb C\setminus\mathbb R$, where
$  p_{n,N}(z) = \kappa_{n,N} z^n + \cdots$,
is the $n$-th degree orthonormal polynomial with respect to the
weight $|x|^{2\alpha}e^{-NV(x)}$, and $\kappa_{n,N}$ is the
leading coefficient of $p_{n,N}$.

The correlation kernel $K_{n,N}$ can be expressed in terms of the solution
of this RH problem. Indeed, using the Christoffel-Darboux formula for
orthogonal polynomials, we get from (\ref{1-kern}), (\ref{RHP Y: solution}),
and the fact that $\det Y \equiv 1$,
\begin{align}\nonumber
    K_{n,N}(x,y) &= |x|^\alpha
    e^{-\frac{1}{2}NV(x)}|y|^\alpha e^{-\frac{1}{2}NV(y)}\frac{\kappa_{n-1,N}}{\kappa_{n,N}}
    \frac{p_{n,N}(x)p_{n-1,N}(y)-p_{n-1,N}(x)p_{n,N}(y)}{x-y} \\[2ex]
    \label{KinY}
    &= |x|^\alpha e^{-\frac{1}{2}NV(x)}|y|^\alpha e^{-\frac{1}{2}NV(y)}
    \frac{1}{2\pi i(x-y)}
    \begin{pmatrix}
        0 & 1
    \end{pmatrix}
    Y_{\pm}^{-1}(y) Y_{\pm}(x)
    \begin{pmatrix}
        1 \\ 0
    \end{pmatrix}.
\end{align}

The asymptotics of $K_{n,N}$ follows from a steepest descent analysis of the RH problem for
$Y$, see \cite{CK,DKMVZ2,DKMVZ1,KV1,KV2,Vanlessen2}. The Deift/Zhou steepest descent analysis
consists of a series of explicit transformations $Y\mapsto T\mapsto S\mapsto R$ so that it
leads to a RH problem for $R$ which is normalized at infinity and which has jumps uniformly
close to the identity matrix $I$. Then $R$ itself is uniformly close to $I$. By going back in
the series of transformations we then have the asymptotics for $Y$ from which the asymptotics
of $K_{n,N}$ in different scaling regimes can be deduced.

The main issue of the present situation is the construction of a local
parametrix near $0$ with the aid of the RH problem for $\Psi_{\alpha}$
introduced in Section \ref{section: Painleve II}. For the case $\alpha=0$ this
was done in \cite{CK} and we use the ideas introduced in that paper.

Throughout the rest of the paper we use the notation
\begin{equation} \label{notation}
    t=n/N, \qquad \mbox{and} \qquad V_t = \frac{1}{t} V.
\end{equation}

\subsection{First transformation $Y\mapsto T$}

In the first transformation we normalize the RH problem at
infinity. The standard approach would be to use the equilibrium
measure in the external field $V_t$, see \cite{Deift,SaTo}.
This is the probability measure  that minimizes
\[ I_{V_t}(\mu) = \iint \log \frac{1}{|x-y|} d\mu(x) d\mu(y)
    +  \int V_t(x) d\mu(x) \]
among all Borel probability measures $\mu$ on $\mathbb R$.
The minimizer for $t=1$ has density $\psi_V$ which by assumption
vanishes at the origin. For $t<1$, the origin is outside
of the support and for $t$ slightly less than $1$,  there is a gap in the
support around $0$. An annoying consequence is that the equality
in the variational conditions is not valid near the origin.
Therefore, a modified measure $\nu_t$ was introduced in
\cite{CK} to overcome this problem.

Here, we follow \cite[Section 3]{CK}. We take a small $\delta_0>0$
so that $\psi_V(x) > 0$ for $x \in [-\delta_0, \delta_0] \setminus \{0\}$,
and we consider the problem to minimize $I_{V_t}(\nu)$ among all
signed measures $\nu=\nu^+-\nu^-$ where $\nu^\pm$ are nonnegative
measures such that $\int d\nu=1$ and $\supp(\nu^-) \subset[-\delta_0,\delta_0]$.
There is a unique minimizer which we denote by $\nu_t$. This signed measure is
absolutely continuous with density $\psi_t$ and its support $S_t = [a_t,b_t]$
is an interval if $t$ is sufficiently close to $1$. The following
variational conditions are satisfied: there exists a constant
$\ell_t\in\mathbb R$ such that
\begin{align}
    \label{variational condition 1}
    & 2\int\log|x-y|\psi_t(y)dy-V_t(x)+\ell_t=0, && \mbox{for $x\in [a_t,b_t]$,} \\
    \label{variational condition 2}
    & 2\int\log|x-y|\psi_t(y)dy-V_t(x)+\ell_t\leq 0, && \mbox{for $x \in \mathbb
    R$.}
\end{align}
In addition, it was shown in \cite{CK} that for $t$ sufficiently close to $1$,
\begin{equation} \label{psitassqrt}
    \psi_t(x) = \frac{1}{\pi} (-Q_t(x))^{1/2}, \qquad \mbox{for } x \in [a_t,b_t],
    \end{equation}
where
\begin{equation} \label{defqt}
    Q_t(z) = \left(\frac{V'(z)}{2t}\right)^2 - \frac{1}{t} \int \frac{V'(z)-V'(y)}{z-y}
    \psi_t(y) dy.
    \end{equation}
For $t > 1$, we take the square root in (\ref{psitassqrt}) which is positive for $x =0$,
while for $t < 1$ we take the square root which is negative for $x=0$.

For the first transformation, we introduce the
following `$g$-function' associated with  $\nu_t$,
\begin{equation}\label{definition: g}
g_t(z)= \int \log(z-y) d\nu_t(y) = \int \log (z-y)\psi_t(y)dy,
\qquad \mbox{for $z\in\mathbb C\setminus\mathbb R$,}
\end{equation}
where we take the branch cut of the logarithm along the negative
real axis.  We define
\begin{equation}\label{TinY}
    T(z)=e^{\frac{1}{2} n\ell_t \sigma_3} Y(z) e^{-ng_t(z)\sigma_3}
    e^{-\frac{1}{2}n\ell_t\sigma_3}, \qquad \mbox{for }
    z\in\mathbb C\setminus\mathbb R.
\end{equation}
We also use the functions
\begin{align} \label{defvarphit}
    \varphi_t(z) & = \int_{b_t}^z (Q_t(s))^{1/2} ds, \\[1ex]
    \label{deftildevarphit}
     \tilde{\varphi}_t(z) & = \int_{a_t}^z (Q_t(s))^{1/2} ds,
\end{align}
where the path of integration does not cross the real axis.
The relations that exist between $g_t$, $\varphi_t$ and $\tilde{\varphi}_t$
are described in \cite[Section 5.2]{CK}. Using these, we find that $T$
is the unique solution of the following RH problem.

\subsubsection*{RH problem for $T$:}
\begin{itemize}
    \item[(a)] $T:\mathbb C\setminus\mathbb R\to\mathbb C^{2\times
        2}$ is analytic.
    \item[(b)] $T_+(x)=T_-(x)v_T(x)$ for $x\in\mathbb R$, with
        \[
            v_T(x) =
            \begin{cases}
                \begin{pmatrix}
                    e^{2n \varphi_{t,+}(x)} & |x|^{2\alpha} \\
                    0 & e^{2n \varphi_{t,-}(x)}
                \end{pmatrix}, & \mbox{for $x\in (a_t,b_t)$,} \\[3ex]
                \begin{pmatrix}
                    1 & |x|^{2\alpha} e^{-2n \varphi_t(x)} \\
                    0 & 1
                \end{pmatrix}, & \mbox{for $x \in (b_t, \infty)$,} \\[3ex]
                \begin{pmatrix}
                    1 & |x|^{2\alpha} e^{-2n \tilde{\varphi}_t(x)} \\
                    0 & 1
                \end{pmatrix}, & \mbox{for $x \in (-\infty, a_t)$.}
            \end{cases}
        \]
    \item[(c)] $T(z)=I+\bigO(1/z)$,\qquad as $z\to\infty$.
    \item[(d)] $T$ has the same behavior as $Y$ near the origin,
    given by (\ref{RHP Y: d}).
\end{itemize}

\subsection{Second transformation $S\mapsto T$}

\begin{figure}[t]
\begin{center}
    \setlength{\unitlength}{1mm}
    \begin{picture}(100,26)(0,11.5)
        \put(0,25){\line(1,0){100}}
        \put(31,25){\thicklines\vector(1,0){.0001}} \put(66,25){\thicklines\vector(1,0){.0001}}
        \put(9,25){\thicklines\vector(1,0){.0001}} \put(94,25){\thicklines\vector(1,0){.0001}}

        \put(15,25){\thicklines\circle*{.8}}
        \put(45,25){\thicklines\circle*{.8}}
        \put(85,25){\thicklines\circle*{.8}}

        \qbezier(15,25)(30,43)(45,25) \put(31,34){\thicklines\vector(1,0){.0001}}
        \qbezier(15,25)(30,7)(45,25) \put(31,16){\thicklines\vector(1,0){.0001}}
        \qbezier(45,25)(65,43)(85,25) \put(66,34){\thicklines\vector(1,0){.0001}}
        \qbezier(45,25)(65,7)(85,25) \put(66,16){\thicklines\vector(1,0){.0001}}

        \put(12.7,27.2){\small $a_t$}
        \put(44.5,27.2){\small 0}
        \put(84.5,27.2){\small $b_t$}
    \end{picture}
    \caption{The lens shaped contour $\Sigma_S$ going through the origin.}
    \label{figure: opening lens}
\end{center}
\end{figure}
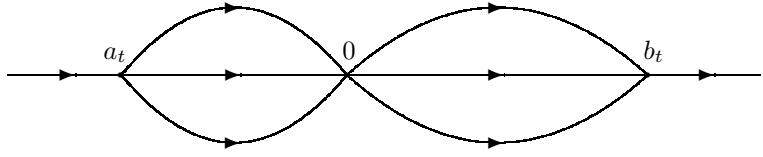

In this subsection, we open the lens as in Figure \ref{figure: opening lens}.
The opening of the lens is based on the factorization of the jump
matrix $v_T$ for $x\in (a_t,b_t)$, which is
\begin{align}
    \nonumber
    v_T(x) &=
    \begin{pmatrix}
        e^{2n \varphi_{t,+}(x)} & |x|^{2\alpha} \\
        0 & e^{2n \varphi_{t,-}(x)}
    \end{pmatrix} \\[1ex]
    \label{factorization}
    & =
    \begin{pmatrix}
        1 & 0 \\
        |x|^{-2\alpha} e^{2n \varphi_{t,-}(x)} & 1
    \end{pmatrix}
    \begin{pmatrix}
        0 & |x|^{2\alpha} \\
        -|x|^{-2\alpha} & 0
    \end{pmatrix}
    \begin{pmatrix}
        1 & 0 \\
        |x|^{-2\alpha}e^{2n\varphi_{t,+}(x)} & 1
    \end{pmatrix}.
\end{align}
We deform the RH problem for $T$ into a RH problem for $S$ by
opening a lens around $[a_t,b_t]$ going through the origin, as
shown in Figure \ref{figure: opening lens}. The precise form of
the lens is not yet specified but for now we choose the lens to be
contained in the region of analyticity of $V$ and we
can do it in such a way that for any given $\delta>0$, there exists
$\gamma>0$ so that, for every $t$ sufficiently close to $1$, we
have that
\begin{equation} \label{inequality: phi}
    \Re \varphi_t(z) < -\gamma,
\end{equation}
for $z$ on the upper and lower lips of the lens with the
exception of $\delta$-neighborhoods of $0$, $a$, and $b$.
See also \cite[Section 5.3]{CK}.

Let $\omega$ be the analytic continuation of $x \mapsto |x|^{2\alpha}$
to $\mathbb C \setminus (i \mathbb R)$, i.e.,
\begin{equation}\label{definition: omega}
    \omega(z)=
    \begin{cases}
        z^{2\alpha},& \mbox{if $\Re z>0$,} \\
        (-z)^{2\alpha},& \mbox{if $\Re z<0$.}
    \end{cases}
\end{equation}
The second transformation is then defined by
\begin{equation}\label{SinT}
    S(z)=
    \begin{cases}
        T(z), & \mbox{for $z$ outside the lens.} \\[1ex]
        T(z)
            \begin{pmatrix}
                1 & 0\\
                -\omega(z)^{-1}e^{2n\varphi_t(z)} & 1
            \end{pmatrix}, & \mbox{for $z$ in the upper parts of the
            lens,}\\[3ex]
        T(z)
            \begin{pmatrix}
                1 & 0 \\
                \omega(z)^{-1}e^{2n\varphi_t(z)}& 1
            \end{pmatrix}, & \mbox{for $z$ in the lower parts of the lens.}
    \end{cases}
\end{equation}
Then $S$ is the unique solution of the following RH problem
posed on the contour $\Sigma_S$ which is the union of $\mathbb R$
with the upper and lower lips of the lens.
\subsubsection*{RH problem for $S$:}
\begin{itemize}
    \item[(a)] $S:\mathbb C\setminus\Sigma_S\to\mathbb C^{2\times 2}$ is analytic.
    \item[(b)] $S_+ = S_- v_S$ on $\Sigma_S$, where
    \[
        v_S(z) =
        \begin{cases}
            \begin{pmatrix}
                    1 & 0 \\
                    \omega(z)^{-1}e^{2n \varphi_t(z)} & 1 \\
                \end{pmatrix}, & \mbox{for $z\in \Sigma_S \setminus \mathbb R$,}
                \\[3ex]
            \begin{pmatrix}
                    0 & |z|^{2\alpha} \\
                    -|z|^{-2\alpha} & 0 \\
                \end{pmatrix}, & \mbox{for $z\in (a_t,b_t)$,}
                \\[3ex]
            \begin{pmatrix}
                    1 & |z|^{2\alpha} e^{-2n \varphi_t(z)} \\
                    0 & 1
                \end{pmatrix}, & \mbox{for $z\in (b_t,\infty)$,}
                \\[3ex]
            \begin{pmatrix}
                    1 & |z|^{2\alpha} e^{-2n \tilde{\varphi}_t(z)} \\
                    0 & 1
                \end{pmatrix}, & \mbox{for $z\in (-\infty, a_t)$.}
        \end{cases}
    \]
    \item[(c)] $S(z)=I+\bigO(1/z)$, \qquad as $z\to \infty$.
    \item[(d)] $S$ has the following behavior near the origin. If
    $\alpha<0$,
        \begin{equation}\label{RHP S: d1}
            S(z)=
            \bigO\begin{pmatrix}
                1 & |z|^{2\alpha} \\
                1 & |z|^{2\alpha}
            \end{pmatrix},\qquad \mbox{as $z\to
            0,z\in\mathbb C\setminus\Sigma_S$,}
        \end{equation}
        and if $\alpha\geq 0$,
        \begin{equation}\label{RHP S: d2}
            S(z)=\begin{cases}
                \bigO\begin{pmatrix}
                    1 & 1 \\
                    1 & 1
                \end{pmatrix}, & \mbox{as $z\to 0$ from outside the
                lens,} \\[3ex]
                \bigO\begin{pmatrix}
                    |z|^{-2\alpha} & 1 \\
                    |z|^{-2\alpha} & 1
                \end{pmatrix}, & \mbox{as $z\to 0$ from inside the lens.}
            \end{cases}
        \end{equation}
\end{itemize}

\subsection{Parametrix away from special points}

On the lips of the lens and on $(-\infty, a_t) \cup (b_t,
\infty)$, the jump matrix for $S$ is close to the identity matrix
if $n$ is large and $t$ is close to $1$. This follows from the
inequality (\ref{inequality: phi}) and the fact that $\varphi_t(x)
> 0$ for $x > b_t$ and $\tilde{\varphi}_t(x) > 0$ for $x < a_t$.
Ignoring these jumps we are led to the following RH problem.

\subsubsection*{RH problem for $P^{(\infty)}$:}
\begin{itemize}
    \item[(a)] $P^{(\infty)}:\mathbb C\setminus [a_t,b_t] \to \mathbb C^{2\times 2}$ is analytic.
    \item[(b)] $P^{(\infty)}_+(x)=P^{(\infty)}_-(x)
            \begin{pmatrix}
                0 & |x|^{2\alpha}\\
                -|x|^{-2\alpha}  & 0
            \end{pmatrix}$,\qquad for $x\in (a_t,b_t) \setminus\{0\}$.
    \item[(c)] $P^{(\infty)}(z)=I+\bigO(1/z)$,\qquad as $z\to\infty$.
\end{itemize}

Note that $P^{(\infty)}$ depends on $n$ and $N$ through the
parameter $t$. As in \cite{Kuijlaars,KMVV,KV2} we construct
$P^{(\infty)}$ in terms of the Szeg\H{o} function $D$ associated
with $|x|^{2\alpha}$ on $(a_t,b_t)$. This is an analytic function
in $\mathbb{C}\setminus[a_t,b_t]$ satisfying
$D_+(x)D_-(x)=|x|^{2\alpha}$ for $x\in(a_t,b_t)\setminus\{0\}$ and
which does not vanish anywhere in $\overline{\mathbb
C}\setminus[a_t,b_t]$. It is easy to check that $D$ is given by
\begin{equation}\label{definition: szego}
    D(z)=z^\alpha\phi\left(\frac{2z-a_t-b_t}{b_t-a_t}\right)^{-\alpha},
    \qquad\mbox{for $z\in\mathbb{C}\setminus[a_t,b_t]$,}
\end{equation}
where $\phi(z)=z+(z-1)^{1/2}(z+1)^{1/2}$ is the conformal map from
$\mathbb{C}\setminus[-1,1]$ onto the exterior of the unit circle.
Since $\phi(z)=2z+\bigO(1/z)$ as $z\to\infty$ we have
\[
    \lim_{z\to\infty}D(z)=\left(\frac{4}{b_t-a_t}\right)^{-\alpha}\equiv D_\infty.
\]
Now the transformed matrix valued function
\begin{equation} \label{defhatPinfty}
    \widehat P^{(\infty)}=D_\infty^{-\sigma_3}P^{(\infty)}D^{\sigma_3}
    \end{equation}
satisfies conditions (a) and (c) of the RH problem and it has the
jump matrix $\left(\begin{smallmatrix}0 & 1
\\ -1 & 0\end{smallmatrix}\right)$ on $(a_t,b_t)$.
The construction of $\widehat P^{(\infty)}$ has been done in
\cite{Deift,DKMVZ2,DKMVZ1}, and leads us to the solution of the RH
problem for $P^{(\infty)}$:
\begin{equation}\label{Pinfty}
    P^{(\infty)}(z) = D_\infty^{\sigma_3}
        \begin{pmatrix}
            \frac{\beta(z)+\beta(z)^{-1}}{2} & \frac{\beta(z)-\beta(z)^{-1}}{2i}
            \\[1ex]
            \frac{\beta(z)-\beta(z)^{-1}}{-2i} & \frac{\beta(z)+\beta(z)^{-1}}{2}
        \end{pmatrix}
        D(z)^{-\sigma_3}, \qquad\mbox{for $z\in\mathbb{C}\setminus[a_t,b_t]$,}
\end{equation}
where
\begin{equation}\label{definition: beta}
    \beta(z)=\frac{(z-b_t)^{1/4}}{(z-a_t)^{1/4}},\qquad\mbox{for $z\in\mathbb{C}\setminus[a_t,b_t]$.}
\end{equation}

\subsection{Parametrix near endpoints}

The jump matrices of $S$ and $P^{(\infty)}$ are not uniformly
close to each other near the origin and near the endpoints of
$[a_t,b_t]$. We surround $a_1$ and $b_1$ (the endpoints of $S_V$)
with small disks $U_{\delta}(a)$ and $U_{\delta}(b)$ of radius
$\delta$. For $t$ sufficiently close to $1$, the endpoints $a_t$
and $b_t$ are in these disks, and then local parametrices
$P^{(a_t)}$ and $P^{(b_t)}$  can be constructed with Airy
functions as in \cite{Deift,DKMVZ2,DKMVZ1}.

\subsection{Parametrix near the origin}

Near the origin a local parametrix will be constructed with the
aid of the RH problem for $\Psi_{\alpha}$ of Section \ref{section: Painleve II}.
Let $U_{\delta}$ be a small disk with center at $0$
and radius $\delta>0$. We seek a $2\times 2$ matrix valued
function $P$ in $U_{\delta}$ with the same jumps as $S$, with the
same behavior as $S$ near the origin, and which matches with
$P^{(\infty)}$ on the boundary $\partial U_{\delta}$ of the disk.
We thus seek a $2\times 2$ matrix valued function that satisfies
the following RH problem.

\subsubsection*{RH problem for \boldmath$P$:}
\begin{itemize}
\item[(a)]
    $P$ is defined and analytic in $U_{\delta'}\setminus\Sigma_S$ for some $\delta' > \delta$.
\item[(b)]
    On $\Sigma_S \cap U_{\delta}$, $P$ satisfies the jump relations
    \begin{align}
        \label{RHP P: b1}
        P_+(z) &=
            P_-(z)
            \begin{pmatrix}
                1 & 0 \\
                \omega(z)^{-1}e^{2n\varphi_t(z)} & 1
            \end{pmatrix},
            && \mbox{for $z\in \left(\Sigma_S \setminus \mathbb R \right) \cap U_{\delta}$,}
        \\[1ex]
        \label{RHP P: b2}
        P_+(x) &=
            P_-(x)
            \begin{pmatrix}
                0 & |x|^{2\alpha} \\
                -|x|^{-2\alpha} & 0
            \end{pmatrix},
            && \mbox{for $x \in (-\delta, \delta) \setminus\{0\}$.}
    \end{align}
\item[(c)] $P$ satisfies the matching condition
    \begin{equation} \label{matchingP}
        P(z) = (I + \bigO(n^{-1/3})) P^{(\infty)}(z);
        \end{equation}
    as $n,N\to\infty$ such that $ n^{2/3} (n/N - 1) \to L$,
    uniformly for $z \in \partial U_{\delta}\setminus\Sigma_S$.
\item[(d)]
    $P$ has the same behavior near the origin as $S$, given by (\ref{RHP S: d1}) and (\ref{RHP S: d2}).
\end{itemize}

In order to solve the RH problem for $P$ we work as follows. First, we seek $P$
such that it satisfies conditions (a), (b), and (d). To do this, we transform
(in the first step) the RH problem into a RH problem for $\widehat P$ with
constant jump matrices. In the second step we solve the RH problem for
$\widehat P$ explicitly by using the RH problem for $\Psi_{\alpha}$. In the
final step we take also the matching condition (c) into account.

\subsubsection*{Step 1: Transformation to constant jump matrices}

In the first step we transform the RH problem for $P$ into a RH
problem for $\widehat P$ with constant jump matrices. We seek $P$
in the form
\begin{align}
    \label{P in Phat: eq1}
    P(z) & = E(z) \widehat P(z) e^{n\varphi_t(z)\sigma_3}
            e^{\frac{1}{2} \pi i \alpha \sigma_3} z^{-\alpha\sigma_3},
            && \mbox{if $\Im z>0$,}
    \\[1ex]
    \label{P in Phat: eq2}
    P(z) & = E(z) \widehat P(z) e^{-n\varphi_t(z)\sigma_3}
            e^{\frac{1}{2} \pi i \alpha \sigma_3}
            \begin{pmatrix}
                0 & -1 \\
                1 & 0
            \end{pmatrix} z^{-\alpha\sigma_3},
            && \mbox{if $\Im z<0$,}
\end{align}
where the invertible matrix valued function $E=E_{n,N}$ (we
suppress notation of the indices) is analytic in  $U_{\delta'}$ and
where the branch cut of $z^\alpha$ is chosen along the negative
real axis.

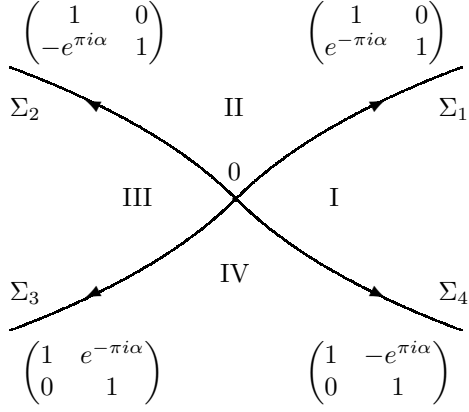
\begin{figure}[t]
\begin{center}
    \setlength{\unitlength}{1truemm}
    \begin{picture}(101.5,58.5)(0,-3.5)
        \put(77,38.5){\small $\Sigma_1$}
        \put(20,38.5){\small $\Sigma_2$}
        \put(20,14){\small $\Sigma_3$}
        \put(77,14){\small $\Sigma_4$}

        \put(62.5,26.5){\small $\I$}
        \put(48.5,38.5){\small $\II$}
        \put(35,26.5){\small $\III$}
        \put(48,16.5){\small $\IV$}

        \put(21.5,3.5){\small
                $\begin{pmatrix}
                    1 & e^{-\pi i\alpha} \\
                    0 & 1
                \end{pmatrix}$ }
        \put(21.5,48.75){\small
                $\begin{pmatrix}
                    1 & 0 \\
                    -e^{\pi i\alpha} & 1
                \end{pmatrix}$ }
        \put(59.25,3.5){\small
                $\begin{pmatrix}
                    1 & -e^{\pi i\alpha} \\
                    0 & 1
                \end{pmatrix}$ }
        \put(59.25,48.75){\small
                $\begin{pmatrix}
                    1 & 0 \\
                    e^{-\pi i\alpha} & 1
                \end{pmatrix}$ }

        \put(50,27.5){\thicklines\circle*{.8}}
        \put(49,30){\small 0}

        \qbezier(50,27.5)(40,17.5)(20,10)
        \qbezier(50,27.5)(60,37.5)(80,45)
        \qbezier(50,27.5)(40,37.5)(20,45)
        \qbezier(50,27.5)(60,17.5)(80,10)

        \put(70,40.75){\thicklines\vector(2,1){.0001}}
        \put(70,14.25){\thicklines\vector(2,-1){.0001}}
        \put(30,40.75){\thicklines\vector(-2,1){.0001}}
        \put(30,14.25){\thicklines\vector(-2,-1){.0001}}
    \end{picture}
    \caption{Contour and jumps for the RH problem for $\widehat{P}$.}
    \label{figure: RHP Phat}
\end{center}
\end{figure}

Using  (\ref{RHP P: b2}), (\ref{P in Phat: eq1}) and
(\ref{P in Phat: eq2}), and keeping track of the branches of $z^{\alpha}$, we
can easily check that $\widehat P$ has no jumps on
$(-\delta,\delta)\setminus\{0\}$. What remains are jumps on the
contour $(\Sigma_S \setminus \mathbb R) \cap U_{\delta} = \bigcup_{j=1}^4\Sigma_j$,
which is shown in Figure \ref{figure: RHP Phat}. We have reversed the orientation of
$\Sigma_2$ and $\Sigma_3$ towards infinity, so that now the
orientation of the $\Sigma_j$'s corresponds to the orientation of
the $\Gamma_j$'s in Figure \ref{figure: RHP Psi}. The contour
divides $U_{\delta}$ into four regions $\I,\II,\III$ and $\IV$, also
shown in Figure \ref{figure: RHP Phat}.

We will now determine the jump relations for $\widehat P$. By
(\ref{definition: omega}), (\ref{RHP P: b1}), and
(\ref{P in Phat: eq1}), $\widehat P$ should have
the following jump matrix on $\Sigma_1$,
\begin{align}
    \nonumber
    \widehat P_-(z)^{-1}\widehat P_+(z) &= e^{n\varphi_t(z)\sigma_3}
        e^{\frac{1}{2}\pi i\alpha\sigma_3} z^{-\alpha\sigma_3}
        \begin{pmatrix}
            1 & 0 \\
            \omega(z)^{-1}e^{2n\varphi_t(z)} & 1
        \end{pmatrix}z^{\alpha\sigma_3}e^{-\frac{1}{2}\pi i\alpha\sigma_3}
            e^{-n\varphi_t(z)\sigma_3}
    \\
    & = \begin{pmatrix}
        1 & 0 \\
        \omega(z)^{-1}z^{2\alpha}e^{-\pi i\alpha} & 1
    \end{pmatrix}
    = \begin{pmatrix}
        1 & 0 \\
        e^{-\pi i\alpha} & 1
    \end{pmatrix}.
\end{align}
For $z\in\Sigma_2$ we have, because of the reversal of the
orientation, an extra minus sign in the  $(2,1)$-entry of the jump matrix. The result is
\begin{equation}
    \widehat P_-(z)^{-1}\widehat P_+(z) = \begin{pmatrix}
        1 & 0 \\
        -\omega(z)^{-1}z^{2\alpha}e^{-\pi i\alpha} & 1
    \end{pmatrix}
    =\begin{pmatrix}
        1 & 0 \\
        - e^{\pi i\alpha} & 1
    \end{pmatrix},
\end{equation}
where the last equality follows from the fact that
$\omega(z)^{-1}z^{2\alpha}=e^{2\pi i\alpha}$ by (\ref{definition: omega}),
 since $\Re z < 0$ in this case. Using
equations (\ref{RHP P: b2}) and (\ref{P in Phat: eq2}), the jump
matrices for $\widehat P$ on $\Sigma_3$ and $\Sigma_4$ can be
determined similarly. The result is that
\begin{equation}
    \widehat P_+(z)=
    \begin{cases}
        \widehat P_-(z)
            \begin{pmatrix}
                1 & e^{-\pi i \alpha} \\
                0 & 1
            \end{pmatrix},
            & \mbox{for $z\in\Sigma_3$,}
            \\[3ex]
        \widehat P_-(z)
            \begin{pmatrix}
                1 & -e^{\pi i \alpha} \\
                0 & 1
            \end{pmatrix},
            & \mbox{for $z\in\Sigma_4$.}
    \end{cases}
\end{equation}

We arrive at the following RH problem for $\widehat P$.
If it is satisfied by $\widehat P$ then $P$
defined by (\ref{P in Phat: eq1})-(\ref{P in Phat: eq2})
satisfies the parts (a), (b), and (d) of the RH problem for $P$.

\subsubsection*{RH problem for $\widehat P$:}

\begin{itemize}
\item[(a)] $\widehat P$ is defined and analytic in $U_{\delta'}\setminus\bigcup_j\Sigma_j$
    for some $\delta'>\delta$.
\item[(b)] $\widehat P$ satisfies the following jump
    relations
    \begin{equation}
        \widehat P_+(z) =
        \begin{cases}
            \widehat P_-(z)
            \begin{pmatrix}
                1 & 0 \\
                e^{-\pi i \alpha} & 1
            \end{pmatrix},
            & \mbox{for $z\in\Sigma_1$,}
            \\[3ex]
            \widehat P_-(z)
            \begin{pmatrix}
                1 & 0 \\
                -e^{\pi i \alpha} & 1
            \end{pmatrix},
            & \mbox{for $z\in\Sigma_2$,}
            \\[3ex]
            \widehat P_-(z)
            \begin{pmatrix}
                1 & e^{-\pi i \alpha} \\
                0 & 1
            \end{pmatrix},
            & \mbox{for $z\in\Sigma_3$,}
            \\[3ex]
            \widehat P_-(z)
            \begin{pmatrix}
                1 & -e^{\pi i \alpha} \\
                0 & 1
            \end{pmatrix},
            & \mbox{for $z\in\Sigma_4$.}
        \end{cases}
    \end{equation}
\item[(c)] $\widehat P$ has the following behavior near the origin. If $\alpha<0$,
    \begin{equation}\label{RHP Phat: d1}
        \widehat P(z)=
        \bigO\begin{pmatrix}
            |z|^\alpha & |z|^\alpha \\
            |z|^\alpha & |z|^\alpha
        \end{pmatrix},
        \qquad \mbox{as $z\to 0$,}
    \end{equation}
    and if $\alpha\geq0$,
    \begin{equation}
        \widehat P(z)=
        \begin{cases}
            \bigO\begin{pmatrix}
                |z|^{-\alpha} & |z|^{-\alpha} \\
                |z|^{-\alpha} & |z|^{-\alpha}
            \end{pmatrix},
                & \mbox{as $z\to 0$, $z\in\I\cup\III$,}
            \\[3ex]
            \bigO\begin{pmatrix}
                |z|^{\alpha} & |z|^{-\alpha} \\
                |z|^{\alpha} & |z|^{-\alpha}
            \end{pmatrix},
                & \mbox{as $z\to 0$, $z\in\II$,}
            \\[3ex]
            \bigO\begin{pmatrix}
                |z|^{-\alpha} & |z|^{\alpha} \\
                |z|^{-\alpha} & |z|^{\alpha}
            \end{pmatrix},
                & \mbox{as $z\to 0$, $z\in\IV$.}
        \end{cases}
    \end{equation}
\end{itemize}
Note that if $\widehat P$ has the behavior near the origin
as described in part (c) of the RH problem, then $P$ defined
by (\ref{P in Phat: eq1}) and (\ref{P in Phat: eq2}) has
the same behavior near the origin as $S$, as required by
part (d) of the RH problem for $P$.

\subsubsection*{Step 2: Construction of $\widehat P$}

Observe that the jump matrices and the behavior near the origin of the RH
problem for $\widehat P$ correspond exactly to the jump matrices and the
behavior near the  origin of the RH problem for $\Psi_{\alpha}$. We use the
solution of the latter RH problem to solve the RH problem for $\widehat P$.

We seek $\widehat P$ in the form
\begin{equation}\label{solution RHP Phat}
    \widehat
    P(z)=\Psi_{\alpha}\left(n^{1/3}f(z);n^{2/3}s_t(z)\right),
\end{equation}
where $f$ and $s_t$ are analytic functions on $U_{\delta}$
which are real on $(-\delta, \delta)$, and $s_t$ is such that
\begin{equation} \label{pole condition}
    n^{2/3} s_t(z) \in \mathbb C \setminus \mathcal P_{\alpha},
    \qquad \mbox{for $z \in U_{\delta}$,}
\end{equation}
where $\mathcal P_{\alpha}$ is the set of poles of $q_{\alpha}$.
In addition, $f$ is a conformal map from $U_{\delta}$ onto a convex neighborhood
$f(U_{\delta})$ of $0$ such that $f(0)=0$ and $f'(0)>0$.
Depending on $f$ we open the lens around $[a_t,b_t]$ such that
that $f(\Sigma_i)=\Gamma_i$ for $i=1,2,3,4$,  where the
$\Gamma_i$'s are the jump contours for the RH problem for $\Psi_{\alpha}$,
see Figure \ref{figure: RHP Psi}. Recall that the lens was not
fully specified and we still have the freedom to make this choice.

It remains to determine $f$ and $s_t$ so that the matching
condition for $P$ is also satisfied. Here we again follow \cite{CK}.
As in \cite[Section 5.6]{CK} we take
\begin{equation}\label{definition: f}
    f(z) =
        \left[\frac{3}{4} \int_0^z (-Q_1(y))^{1/2} dy \right]^{1/3}
        =z\left(\frac{\pi\psi_V''(0)}{8}\right)^{1/3}+\bigO\left(z^2\right),\qquad \mbox{as $z\to 0$,}
\end{equation}
and
\begin{equation}\label{definition: st}
    s_t(z) f(z)= \int_0^z \left( (-Q_t(y))^{1/2} - (-Q_1(y))^{1/2} \right) dy.
\end{equation}
Then $f$ is analytic with $f(0) =0$ and $f'(0) > 0$, it does not depend on $t$,
and it is a conformal mapping on $U_{\delta}$ provided $\delta$ is small
enough. Since the right-hand side of (\ref{definition: st}) is analytic and
vanishes for $z = 0$, we can divide by $f(z)$ and obtain an analytic function $s_t$.
From \cite[(5.29)]{CK}, we get that there exists a constant $K>0$ such that
\begin{equation}\label{st}
|s_t(z)-\pi c^{1/3}(t-1)w_{S_V}(0)|\leq K(t-1)|z|+o(t-1)\qquad \textrm{ as }t\to 1,
\end{equation}
uniformly for $z$ in a neighborhood of $0$.
Now assume that $|n^{2/3}(t-1)|\leq M$ and $n$ large enough.
Then it easily follows from (\ref{st}) and
the fact that $q_\alpha$ has no real poles, that there exists a $\delta>0$,
depending only on $M$, such that
\begin{equation}\label{Ims}
|\Im n^{2/3}s_t(z)|<\min\{|\Im s| \mid s \textrm{ is a pole of }q_\alpha\}\qquad
\textrm{ for }|z|\leq \delta.
\end{equation}
Then (\ref{pole condition})
holds and (\ref{solution RHP Phat}) is well-defined and
analytic since  $\Psi_{\alpha}(\zeta;s)$ is jointly analytic in its two
arguments, see Remark \ref{remark: analyticitypsi}.

It follows from (\ref{definition: f}) and (\ref{definition: st})
that
\begin{equation}\label{condition f and st}
    -i\left[\frac{4}{3} f(z)^3 + s_t(z)f(z)\right]
    =
    \begin{cases}
        \varphi_{t,+}(0) - \varphi_t(z), & \mbox{if $\Im z>0$,} \\[1ex]
        \varphi_{t,+}(0) + \varphi_t(z), & \mbox{if $\Im z<0$,}
    \end{cases}
\end{equation}
see also \cite[Section 5.6]{CK}. Hence  by (\ref{RHP Psi: c}), which by Remark
\ref{remark: analyticitypsi} holds uniformly for $s$ in compact subsets of
$\mathbb C\setminus \mathcal{P}_{\alpha}$, we have
\begin{multline} \label{asymptotics Phat}
    \widehat P(z)
        =
        \Psi_{\alpha}\left(n^{1/3}f(z);n^{2/3}s_t(z)\right)
\left[I+\bigO(1/n^{1/3})\right]e^{n\varphi_{t,+}(0)\sigma_3}
    \\[1ex]\times
        \begin{cases}
            e^{-n\varphi_t(z)\sigma_3}, & \mbox{if $\Im z>0$,} \\[1ex]
            e^{n\varphi_t(z)\sigma_3}, & \mbox{if $\Im z<0$,}
        \end{cases}
\end{multline}
as $n,N\to\infty$, uniformly for $z\in\partial U_{\delta}$.

\subsubsection*{Step 3: Matching condition}

In the final step we determine $E$ such that the matching
condition (c) of the RH problem for $P$ is satisfied. By
(\ref{P in Phat: eq1}), (\ref{P in Phat: eq2}), and
(\ref{asymptotics Phat}) we have for $z \in \partial U_{\delta}$,
\[
    P(z)=
    \begin{cases}
        E(z)\left[I+\bigO(1/n^{1/3})\right]e^{n\varphi_{t,+}(0)\sigma_3}
            e^{\frac{1}{2}\pi i\alpha\sigma_3}z^{-\alpha
            \sigma_3},& \mbox{if $\Im z>0$,}\\[1ex]
        E(z)\left[I+\bigO(1/n^{1/3})\right]e^{n\varphi_{t,+}(0)\sigma_3}
            e^{\frac{1}{2}\pi i\alpha\sigma_3}
            \begin{pmatrix}
                0 & -1 \\
                1 & 0
            \end{pmatrix}
            z^{-\alpha\sigma_3},& \mbox{if $\Im z<0$,}
    \end{cases}
\]
as $n,N\to\infty$. This has to match the outside parametrix
$P^{(\infty)}$, so that  we are led to the following
definition for the prefactor $E(z)$, for $z \in U_{\delta}$,
\begin{equation}\label{definition: E}
    E(z)=
    \begin{cases}
        P^{(\infty)}(z)z^{\alpha\sigma_3}e^{-\frac{1}{2}\pi i\alpha\sigma_3}
            e^{-n\varphi_{t,+}(0)\sigma_3}, & \mbox{if $\Im z>0$,} \\[1ex]
        P^{(\infty)}(z)z^{\alpha\sigma_3}
        \begin{pmatrix}
            0 & 1 \\
            -1 & 0
        \end{pmatrix}
            e^{-\frac{1}{2}\pi i\alpha\sigma_3}
            e^{-n\varphi_{t,+}(0)\sigma_3}, & \mbox{if $\Im z<0$.}
    \end{cases}
\end{equation}
One can check as in \cite{KV2,Vanlessen} that $E$ is
invertible and analytic in a full neighborhood of $U_{\delta}$. In addition
we have the matching condition (\ref{matchingP}).
This completes the construction of the parametrix near the origin.

\subsection{Third transformation: $S\mapsto R$}

\begin{figure}[t]
\begin{center}
    \setlength{\unitlength}{1mm}
    \begin{picture}(106,26)(-2.5,11.5)
        \put(15,25){\thicklines\circle*{.8}} \put(15,25){\circle{15}}
            \put(13.8,31.9){\thicklines\vector(-1,0){.0001}}
        \put(45,25){\thicklines\circle*{.8}} \put(45,25){\circle{15}}
            \put(43.8,31.9){\thicklines\vector(-1,0){.0001}}
        \put(85,25){\thicklines\circle*{.8}} \put(85,25){\circle{15}}
            \put(83.8,31.9){\thicklines\vector(-1,0){.0001}}
        \put(-2.5,25){\line(1,0){10.5}} \put(4.5,25){\thicklines\vector(1,0){.0001}}
        \put(92,25){\line(1,0){10.5}} \put(99,25){\thicklines\vector(1,0){.0001}}

        \qbezier(19,30.78)(30,41)(41,30.78) \put(31,35.85){\thicklines\vector(1,0){.0001}}
        \qbezier(19,19.22)(30,9)(41,19.22) \put(31,14.15){\thicklines\vector(1,0){.0001}}
        \qbezier(49,30.78)(65,41)(81,30.78) \put(66,35.85){\thicklines\vector(1,0){.0001}}
        \qbezier(49,19.22)(65,9)(81,19.22) \put(66,14.15){\thicklines\vector(1,0){.0001}}

        \put(14,27){\small $a_t$}
        \put(44.3,27){\small 0}
        \put(84,27){\small $b_t$}
    \end{picture}
    \caption{The contour $\Sigma_R$ after the third and final
        transformation.}
    \label{figure: system contours R}
\end{center}
\end{figure}
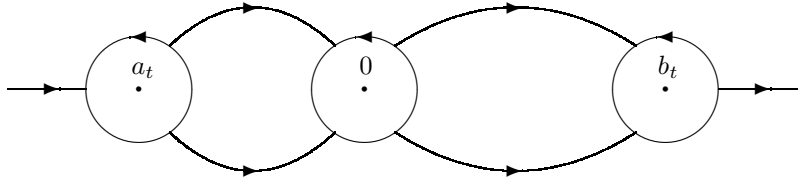

Having the parametrices $P^{(\infty)}$, $P^{(a_t)}$, $P^{(b_t)}$, and $P$, we
now define
\begin{equation}\label{definition: R}
    R(z)=
    \begin{cases}
        S(z)P^{-1}(z), & \mbox{for $z\in U_{\delta}$,}\\[1ex]
        S(z)\left(P^{(a_t)} \right)^{-1}(z), & \mbox{for $z\in U_{\delta}(a)$,}\\[1ex]
        S(z)\left(P^{(b_t)}\right)^{-1}(z), & \mbox{for $z\in U_{\delta}(b)$,}\\[1ex]
        S(z)\left(P^{(\infty)}\right)^{-1}(z), &
            \mbox{for $z\in \mathbb C\setminus
                \left(U_{\delta}\cup U_{\delta}(a) \cup U_{\delta}(b) \cup \Sigma_S\right)$.}
    \end{cases}
\end{equation}
Then $R$ has only jumps on the reduced system of contours $\Sigma_R$
shown in Figure \ref{figure: system contours R}, and
$R$ satisfies the following RH problem, cf.~\cite{CK}.
The circles around $0$, $a_t$ and $b_t$ are oriented counterclockwise.

\subsubsection*{RH problem for $R$:}
\begin{itemize}
    \item[(a)] $R:\mathbb C\setminus\Sigma_R\to\mathbb C^{2\times 2}$ is analytic.
    \item[(b)] $R_+(z)=R_-(z)v_R(z)$ for $z\in \Sigma_R$, with
        \begin{equation} \label{vRdef}
        v_R=
        \begin{cases}
            P^{(\infty)} (P^{(a_t)})^{-1}, & \mbox{on } \partial U_{\delta}(a), \\[1ex]
            P^{(\infty)} (P^{(b_t)})^{-1}, & \mbox{on } \partial U_{\delta}(b), \\[1ex]
            P^{(\infty)} P^{-1}, & \mbox{on }  \partial U_{\delta}, \\[1ex]
            P^{(\infty)} v_S (P^{(\infty)})^{-1}, & \mbox{on the rest of } \Sigma_R.
        \end{cases}
        \end{equation}
    \item[(c)]
        $R(z)=I+\bigO(1/z)$,\qquad as $z\to \infty$,
    \item[(d)] $R$ remains bounded near the intersection points of $\Sigma_R$.
\end{itemize}

Now we let $n, N \to \infty$ such that $|n^{2/3}(n/N-1)|\leq M$, so that
 $\delta$ does not depend on $n$.
Then it follows from the construction of the parametrices that
\begin{equation} \label{vRasymp}
        v_R=
        \begin{cases}
            I + \bigO(1/n), & \mbox{on } \partial U_{\delta}(a)\cup\partial U_{\delta}(b), \\[1ex]
            I + \bigO(n^{-1/3}), & \mbox{on } \partial U_{\delta}, \\[1ex]
            I + \bigO(e^{-\gamma n}), & \mbox{on the rest of } \Sigma_R,
        \end{cases}
\end{equation}
where $\gamma > 0$ is some fixed constant. All $\bigO$-terms hold uniformly on their
respective contours.

For large $n$, the jump matrix of $R$ is close
to the identity matrix, both in $L^{\infty}$ and in $L^2$-sense on $\Sigma_R$.
Then arguments as in \cite{Deift,DKMVZ2,DKMVZ1} (which are based on estimates on
Cauchy operators as well as on contour deformations), guarantee that
\begin{equation} \label{asymp: R}
    R(z)=I+\bigO(n^{-1/3}), \qquad \mbox{uniformly for }z\in \mathbb C \setminus \Sigma_R,
\end{equation}
as $n, N \to \infty$ such that $|n^{2/3}(n/N -1 )| \leq M$.

This completes the
steepest descent analysis. Following the effect of the transformation on the
correlation kernel $K_{n,N}$ and using (\ref{asymp: R}) we will prove the main
Theorem \ref{theorem: main result}. This will be done in the next section.
For the proof of Theorem \ref{theorem: recursie} we need to
expand $v_R(z)$ in (\ref{vRasymp}) up to order $n^{-1/3}$,
from which it follows that
\[
    R(z) = I + \frac{R^{(1)}(z)}{n^{1/3}} + \bigO(n^{-2/3}),
        \qquad \mbox{uniformly for $z \in \mathbb C \setminus \Sigma_R$,}
\]
with an explicitly computable $R^{(1)}(z)$. The asymptotic behavior of the
recurrence coefficients is expressed in terms of $R^{(1)}$ and this leads to
the proof of Theorem \ref{theorem: recursie}. This will be done in Section
\ref{proof recursie}.

\begin{remark}\label{multiinterval}
The steepest descent analysis was done under the assumption that
$\supp(\psi_V)$ consists of one interval. In the multi-interval case, the
construction of the outside parametrix $P^{(\infty)}$ is more complicated,
since it uses $\Theta$-functions as in \cite[Lemma 4.3]{DKMVZ2} and the
Szeg\H{o} function for multiple intervals as in \cite[Section 4]{KV2}. With
these modifications the asymptotic analysis can be carried through in the
multi-interval case without any additional difficulty.
\end{remark}

\section{Proof of Theorem \ref{theorem: main result}}
\label{section: proofmainresult}
As in the statement of Theorem \ref{theorem: main result}, we assume that $n,N\to\infty$ with
$n^{2/3}(t-1)\to L$, where $t=n/N$. Let $M>|L|$ and take $n$ sufficiently large so that
$|n^{2/3}(t-1)|\leq M$. Let $\delta>0$ be such that (\ref{Ims}) holds.
We start by writing the kernel $K_{n,N}$ explicitly in terms of the matrix valued function
$\Phi_{\alpha}$ defined in (\ref{definition: Phi_alpha}). For
notational convenience we introduce
\begin{equation}\label{definition: B}
    B(z)=R(z)E(z),
\end{equation}
where $E$ and $R$ are given by (\ref{definition: E}) and (\ref{definition: R}),
respectively.

\begin{proposition}\label{proposition: KinPhi}
    Let $x,y\in (-\delta,\delta)\setminus\{0\}$. Then
    \begin{multline}\label{KinPhi}
        K_{n,N}(x,y) =
                \frac{1}{2\pi i(x-y)}e^{\frac{1}{2}\pi i\alpha(\sgn(x)+\sgn(y))}
                \begin{pmatrix}
                    0 & 1
                \end{pmatrix}
                \Phi_{\alpha}^{-1}\left(n^{1/3}f(y); n^{2/3}s_t(y)\right)
        \\[1ex]
                \times B^{-1}(y)B(x)\Phi_{\alpha}\left(n^{1/3}f(x); n^{2/3}s_t(x)\right)
                \begin{pmatrix}
                    1 \\
                    0
                \end{pmatrix},
    \end{multline}
    where $\Phi_{\alpha}$ is given by {\rm (\ref{definition: Phi_alpha})}.
\end{proposition}

\begin{proof}
From (\ref{KinY}), (\ref{TinY}), and the fact that $NV=nV_t$, the
kernel $K_{n,N}$ can be written as
\begin{multline*}
    K_{n,N}(x,y) =
    |x|^{\alpha}e^{\frac{1}{2}n\left(2g_{t,+}(x)-V_t(x)+\ell_t\right)}
    |y|^{\alpha}e^{\frac{1}{2}n
    \left(2g_{t,+}(y)-V_t(y)+\ell_t\right)}
    \\ \times\, \frac{1}{2\pi i(x-y)}\
    \begin{pmatrix}
        0 & 1
    \end{pmatrix}
    T_+^{-1}(y)T_+(x)
    \begin{pmatrix}
        1 \\
        0
    \end{pmatrix}.
\end{multline*}
Using the relation
\[
    2g_{t,+}-V_t+\ell_t  = -2\varphi_{t,+} \qquad \mbox{on $[a_t,b_t]$,}
\]
see \cite{CK}, and (\ref{SinT}) to express $T$ in terms of $S$,
we find for $x$ and $y$ in $(a_t,b_t) \setminus \{0\}$,
\begin{align}\label{KinS}
    \nonumber
    K_{n,N}(x,y)
        &= \frac{|x|^\alpha e^{-n\varphi_{t,+}(x)}|y|^\alpha
        e^{-n\varphi_{t,+}(y)}}{2\pi i(x-y)}
        \begin{pmatrix}
            0 & 1
        \end{pmatrix}
        \begin{pmatrix}
            1 & 0 \\
            -|y|^{-2\alpha}e^{2n\varphi_{t,+}(y)} & 1
        \end{pmatrix} S_+^{-1}(y)
        \\[1ex]
        \nonumber
        &\qquad\qquad\qquad\qquad\qquad \times S_+(x) \begin{pmatrix}
            1 & 0 \\
            |x|^{-2\alpha}e^{2n\varphi_{t,+}(x)} & 1
        \end{pmatrix}
        \begin{pmatrix}
            1 \\
            0
        \end{pmatrix}
        \\[3ex]
        \nonumber
        &=
        \frac{1}{2\pi i(x-y)}
        \begin{pmatrix}
            -1 & 1
        \end{pmatrix}
        |y|^{-\alpha\sigma_3}e^{n\varphi_{t,+}(y)\sigma_3}S^{-1}_+(y)
        \\[1ex]
        & \qquad\qquad\qquad\qquad\qquad \times
        S_+(x)|x|^{\alpha\sigma_3}e^{-n\varphi_{t,+}(x)\sigma_3}
        \begin{pmatrix}
            1 \\ 1
        \end{pmatrix}.
\end{align}

We further simplify this expression by writing $S$ in
terms of $R$ and the parametrix $P$ near the origin. Consider the
case that $x\in(0,\delta)$. Then, since $S_+(x)=R(x) P_+(x)$ by (\ref{definition: R}), we
have by (\ref{P in Phat: eq1}),
\begin{equation} \label{hulp0}
    S_+(x) = B(x)\widehat P(x) e^{\frac{1}{2}\pi i\alpha\sigma_3}
    e^{n\varphi_{t,+}(x)\sigma_3}|x|^{-\alpha\sigma_3},\qquad
    \mbox{for $x\in(0,\delta)$,}
\end{equation}
where  $B$ is given by (\ref{definition: B}). By (\ref{hulp0}), (\ref{solution
RHP Phat}), and (\ref{definition: Phi_alpha}) we then find for
$x\in(0,\delta)$,
\begin{align}
    \nonumber
    & S_+(x)|x|^{\alpha\sigma_3}e^{-n\varphi_{t,+}(x)\sigma_3}
        \begin{pmatrix}
            1 \\ 1
        \end{pmatrix}
    \\[1ex]
    \nonumber
    &\qquad\qquad = B(x)\Phi_{\alpha}\left(n^{1/3}f(x);n^{2/3}s_t(x)\right)
    \begin{pmatrix}
        1 & 0 \\
        -e^{-\pi i\alpha} & 1
    \end{pmatrix}e^{\frac{1}{2}\pi i\alpha \sigma_3}\begin{pmatrix} 1 \\ 1
    \end{pmatrix} \\[2ex] \label{hulp1}
    &\qquad\qquad = e^{\frac{1}{2}\pi i\alpha\sgn(x)}B(x)
    \Phi_{\alpha}\left(n^{1/3}f(x);n^{2/3}s_t(x)\right)
    \begin{pmatrix} 1 \\ 0 \end{pmatrix}.
\end{align}
A similar calculation shows that (\ref{hulp1}) also holds for
$x \in(-\delta,0)$. Similarly, we have
\begin{multline}
    \begin{pmatrix}
        -1 & 1
    \end{pmatrix}
        |y|^{-\alpha\sigma_3}e^{n\varphi_{t,+}(y)\sigma_3}S^{-1}_+(y)
        \\[2ex] \label{hulp2}
    = e^{\frac{1}{2}\pi i\alpha\sgn(y)}
    \begin{pmatrix}
        0 & 1
    \end{pmatrix}\Phi_{\alpha}^{-1}\left(n^{1/3}f(y);n^{2/3}s_t(y)\right)B^{-1}(y),
\end{multline}
for $y\in(-\delta,\delta)\setminus\{0\}$.
Inserting (\ref{hulp1}) and (\ref{hulp2}) into (\ref{KinS}), we arrive at
(\ref{KinPhi}), which proves the proposition.
\end{proof}

\medskip

\begin{varproof} \textbf{of Theorem \ref{theorem: main result}.}
Let $u,v\in\mathbb R\setminus\{0\}$, and put
$u_n=u/(cn^{1/3})$ and $v_n=v/(cn^{1/3})$ with $c$
given by (\ref{definition: c}). Note that, by (\ref{definition: f}),
\begin{equation}\label{prooftheorem: eq1}
\lim_{n\to \infty}n^{1/3}f(u_n)=u, \qquad
\lim_{n\to \infty}n^{1/3}f(v_n)=v.
\end{equation}
Furthermore, by (\ref{st}), (\ref{definition: c}), and (\ref{definition: s}),
\[|n^{2/3}s_t(z)-s|\leq Kn^{2/3}(t-1)|z|+n^{2/3}o(t-1)+|n^{2/3}(t-1)-L|\pi c^{1/3}w_{S_V}(0)\]
uniformly for $z$ in a neighborhood of $0$. Then it easily follows that, since $n^{2/3}(t-1)\to L$,
\begin{equation}\label{prooftheorem: eq2}
\lim_{n,N\to\infty}n^{2/3}s_t(u_n)=\lim_{n,N\to\infty}n^{2/3}s_t(v_n)=s.
\end{equation}
Now, similar as in \cite{KV2}, we use the fact that the
entries of $B$ are analytic and uniformly bounded in $U_{\delta}$,
to obtain
\begin{equation}\label{prooftheorem: eq3}
\lim_{n,N\to \infty}B^{-1}(v_n)B(u_n)=I.
\end{equation}
Inserting (\ref{prooftheorem: eq1}), (\ref{prooftheorem: eq2}), and
(\ref{prooftheorem: eq3}) into (\ref{KinPhi}), we find that
\begin{align*}
            & \lim_{n,N\to\infty}\frac{1}{cn^{1/3}}K_{n,N}(u_n,v_n)
            \\[1ex]
            &\qquad\qquad= \frac{1}{2\pi i(u-v)}e^{\frac{1}{2}\pi i\alpha(\sgn(u)+\sgn(v))}
                \begin{pmatrix}
                        0 & 1
                \end{pmatrix}
                \Phi_{\alpha}^{-1}(v;s)
                \Phi_{\alpha}(u;s)
                \begin{pmatrix}
                    1 \\ 0
                \end{pmatrix}
            \\[2ex]
            &\qquad\qquad=-e^{\frac{1}{2}\pi i\alpha(\sgn(u)+\sgn(v))}
            \frac{\Phi_{\alpha,1}(u;s)\Phi_{\alpha,2}(v;s)-\Phi_{\alpha,1}(v;s)\Phi_{\alpha,2}(u;s)}{2\pi
            i(u-v)}.
    \end{align*}
This completes the proof of Theorem \ref{theorem: main result}.
\end{varproof}

\section{Proof of Theorem \ref{theorem: recursie}}
\label{proof recursie}

In this section we will determine the asymptotic behavior of the recurrence coefficients
$a_{n,N}$ and $b_{n,N}$ as $n,N \to \infty$ such that $|n^{2/3}(n/N-1)|\leq M$ for some $M>0$.
As in Theorem \ref{theorem: recursie} we assume that $S_V = [a,b]$ is an interval, and that there are no
other singular points besides $0$. Then it follows that $\supp (\psi_t)$ consists of one
interval $[a_t,b_t]$ if $t$ is sufficiently close to $1$. In addition we have that the
endpoints $a_t$ and $b_t$ are real analytic functions in $t$, see \cite[Theorem 1.3]{KM}, so
that
\begin{equation} \label{abanalytic}
a_t=a +\bigO(n^{-2/3}), \qquad b_t = b+\bigO(n^{-2/3}),\end{equation} since $t= n/N =
1+\bigO(n^{-2/3})$.

We make use of the following result, see for example
\cite{Deift,DKMVZ1}. Let $Y$ be the unique solution of the RH
problem for $Y$. There exist $2\times 2$ constant (independent of
$z$, but depending on $n,N$) matrices $Y_1,Y_2$ such that
\[
    Y(z)\begin{pmatrix}
        z^{-n} & 0 \\
        0 & z^n
    \end{pmatrix}
    =I+\frac{Y_1}{z}+\frac{Y_2}{z^2}+\bigO(1/z^3),\qquad\mbox{as
    $z\to\infty$,}
\]
and
\begin{equation}\label{recurrence in Y}
    a_{n,N}=\sqrt{(Y_1)_{12}(Y_1)_{21}}, \qquad
    b_{n,N}=(Y_1)_{11}+\frac{(Y_2)_{12}}{(Y_1)_{12}}.
\end{equation}
We need to determine the constant matrices $Y_1$ and $Y_2$. For
large $|z|$ we have by (\ref{TinY}), (\ref{SinT}) and
(\ref{definition: R}) that
\begin{equation}\label{Yinfty}
    Y(z)=e^{-\frac{1}{2}n\ell_t\sigma_3} R(z)
    P^{(\infty)}(z)e^{ng_t(z)\sigma_3}e^{\frac{1}{2}n\ell_t\sigma_3}.
\end{equation}
So in order to compute $Y_1$ and $Y_2$ we need the asymptotic
behavior of $P^{(\infty)}(z)$, $e^{ng_t(z)\sigma_3}$ and $R(z)$ as
$z\to\infty$.

\subsubsection*{Asymptotic behavior of $P^{(\infty)}(z)$ as $z\to\infty$:}

From (\ref{definition: szego}) and
(\ref{definition: beta}) it is straightforward to determine the
asymptotic behavior of the scalar functions $D(z)$ and $\beta(z)$
as $z\to\infty$. Indeed, as $z \to \infty$,
\begin{multline*}
    \begin{pmatrix}
        \frac{\beta(z)+\beta(z)^{-1}}{2} & \frac{\beta(z)-\beta(z)^{-1}}{2i} \\[1ex]
        \frac{\beta(z)-\beta(z)^{-1}}{-2i} & \frac{\beta(z)+\beta(z)^{-1}}{2}
    \end{pmatrix} \\[2ex]
    = I - \frac{1}{4}(b_t-a_t)\begin{pmatrix}
        0 & -i \\
        i & 0
    \end{pmatrix}\frac{1}{z}
    +\frac{i}{8}(b_t^2-a_t^2)\begin{pmatrix}
        * & 1\\
        -1 & *
    \end{pmatrix}\frac{1}{z^2}+\bigO(1/z^3),
\end{multline*}
and
\[
    D(z)^{-\sigma_3}=
    \left[I-\frac{\alpha}{2}(b_t+a_t)
    \begin{pmatrix}
        1 & 0 \\
        0 & -1
    \end{pmatrix}\frac{1}{z}
    +\begin{pmatrix}
        * & 0 \\
        0 & *
    \end{pmatrix}\frac{1}{z^2}+\bigO(1/z^3)\right]
    D_\infty^{-\sigma_3},
\]
where $*$ denotes an unspecified unimportant entry.
Inserting these equations into (\ref{Pinfty}) and
using (\ref{abanalytic}) gives us the asymptotic behavior of
$P^{(\infty)}$ at infinity,
\begin{equation}\label{Pinfty-infty}
    P^{(\infty)}(z)=I+\frac{P_1^{(\infty)}}{z}+\frac{P_2^{(\infty)}}{z^2}+\bigO(1/z^3),\qquad\mbox{as
    $z\to\infty$,}
\end{equation}
with
\begin{align}
     \nonumber
    P_1^{(\infty)}& =D_\infty^{\sigma_3}
        \begin{pmatrix}
            -\frac{\alpha}{2}(b_t+a_t) & \frac{i}{4}(b_t-a_t)
            \\[1ex]
            -\frac{i}{4}(b_t-a_t) & \frac{\alpha}{2}(b_t+a_t)
        \end{pmatrix}D_\infty^{-\sigma_3} \\[1ex]
    \label{definition: P1infty}
    &=D_\infty^{\sigma_3}
        \begin{pmatrix}
            -\frac{\alpha}{2}(b+a) & \frac{i}{4}(b-a)
            \\[1ex]
            -\frac{i}{4}(b-a) & \frac{\alpha}{2}(b+a)
        \end{pmatrix}D_\infty^{-\sigma_3} + \bigO(n^{-2/3}),
\end{align}
and
\begin{align}
    \nonumber
    P_2^{(\infty)}
    &=
        D_\infty^{\sigma_3}
        \begin{pmatrix}
            * & \frac{i}{8}(\alpha+1)(b_t^2-a_t^2) \\[1ex]
            \frac{i}{8}(\alpha-1)(b_t^2-a_t^2) & *
        \end{pmatrix}
        D_\infty^{-\sigma_3} \\[1ex]
        &=  \label{definition: P2infty}
     D_\infty^{\sigma_3}
        \begin{pmatrix}
            * & \frac{i}{8}(\alpha+1)(b^2-a^2) \\[1ex]
            \frac{i}{8}(\alpha-1)(b^2-a^2) & *
        \end{pmatrix}
        D_\infty^{-\sigma_3} + \bigO(n^{-2/3}).
\end{align}

\subsubsection*{Asymptotic behavior of $e^{ng_t(z)\sigma_3}$ as $z\to\infty$:}

By (\ref{definition: g}) we have
\begin{equation}\label{ginfty}
    e^{ng_t(z)\sigma_3}\begin{pmatrix}
        z^{-n} & 0 \\
        0 & z^n
    \end{pmatrix}=I+\frac{G_1}{z}+\frac{G_2}{z^2}+\bigO(1/z^3),\qquad\mbox{as
    $z\to\infty$,}
\end{equation}
with
\begin{equation} \label{G1def}
    G_1= -n\int_{a_t}^{b_t}y\psi_t(y)dy
    \begin{pmatrix} 1 & 0 \\ 0 & -1 \end{pmatrix},
    \qquad G_2=\begin{pmatrix}
        * & 0 \\
        0 & *
    \end{pmatrix}.
\end{equation}

\subsubsection*{Asymptotic behavior of $R(z)$ as $z\to\infty$:}
The computation of $R_1$ and $R_2$ is more involved.
For $z \in \partial U_{\delta} \cap \mathbb C_+$, we have
by (\ref{vRdef}), (\ref{P in Phat: eq1}), (\ref{solution RHP Phat}),
and (\ref{definition: E}),
\begin{align}
    \nonumber
    v_R(z)
        & = P^{(\infty)}(z) P^{-1}(z)
    \\
    \nonumber
        & = P^{(\infty)}(z) z^{\alpha \sigma_3} e^{-\frac{1}{2} \pi i \alpha \sigma_3}
        e^{-n \varphi_t(z) \sigma_3}
    \Psi_{\alpha}^{-1}(n^{1/3}f(z);n^{2/3}s_t(z)) \\
    & \qquad\qquad\qquad \times
    e^{n \varphi_{t,+}(0) \sigma_3} e^{\frac{1}{2}\pi i \alpha \sigma_3}
    z^{-\alpha \sigma_3} (P^{(\infty)})^{-1}(z).
\end{align}
Using (\ref{Psiinfty}) and (\ref{condition f and st}), we then find
\begin{equation}\label{vRn}
v_R(z)=I+\frac{\Delta^{(1)}(z)}{n^{1/3}}+\bigO(n^{-2/3}),
\end{equation}
where
\begin{multline} \label{defDelta1}
    \Delta^{(1)}(z) =
        - \frac{1}{2i f(z)} P^{(\infty)}(z) z^{\alpha \sigma_3} e^{-\frac{1}{2}\pi i\alpha\sigma_3}
        e^{-n \varphi_{t,+}(0)\sigma_3}
    \\[1ex]
    \times
        \begin{pmatrix}
            u_{\alpha}(n^{2/3} s_t(z)) & q_{\alpha}(n^{2/3} s_t(z)) \\[1ex]
            -q_{\alpha}(n^{2/3}s_t(z)) & - u_{\alpha}(n^{2/3} s_t(z))
        \end{pmatrix}
        e^{n\varphi_{t,+}(0)\sigma_3} e^{\frac{1}{2}\pi i \alpha \sigma_3} z^{-\alpha \sigma_3}
        (P^{(\infty)})^{-1}(z),
\end{multline}
for $z \in \partial U_{\delta} \cap \mathbb C_+$. A similar calculation
leads to an analogous formula for
$z \in \partial U_{\delta} \cap \mathbb C_-$, which together with (\ref{defDelta1})
shows that $\Delta^{(1)}$ has an extension to an analytic function
in a punctured neighborhood of $0$ with a simple pole at $0$.

To calculate the residue at $0$, we use (\ref{definition: szego})
together with the fact $\phi_+(x)=\exp(i\arccos x)$ for
$x\in[-1,1]$ to find
\[ \lim_{z \to 0+i0} \frac{D(z)}{z^{\alpha}} =
    \exp\left(-i \alpha \arccos \left(- \frac{b_t + a_t}{b_t-a_t}\right)\right), \]
so that by (\ref{defhatPinfty})
\begin{equation} \label{limPinfty}
    \lim_{z \to 0+i0} P^{(\infty)}(z) z^{\alpha \sigma_3} e^{-\frac{1}{2}\pi i \alpha \sigma_3}
    = D_{\infty}^{\sigma_3} \widehat P_+^{(\infty)}(0) e^{i\alpha \theta_t\sigma_3},
    \qquad \mbox{with $\theta_t = \arcsin \frac{b_t+a_t}{b_t-a_t}$.}
\end{equation}
Also note that by (\ref{psitassqrt}), (\ref{defvarphit}), (\ref{definition: omegat}) and
\begin{equation} \label{limvarphi}
    -\varphi_{t,+}(0) = \pi i \int_0^{b_t} \psi_t(x) dx = \pi i \omega_t.
\end{equation}
Now use (\ref{definition: f}), (\ref{definition: c}), (\ref{limPinfty}), and
(\ref{limvarphi}) in (\ref{defDelta1}) to find
\begin{multline} \label{resDelta1}
    \textrm{Res} (\Delta^{(1)};0) =
        -\frac{1}{2ic}D_\infty^{\sigma_3}
        \widehat P_+^{(\infty)}(0)e^{i (\pi n\omega_t+\alpha\theta_t) \sigma_3}
    \\[1ex]
    \times
        \begin{pmatrix}
            u_{\alpha}(n^{2/3} s_t(0)) & q_{\alpha}(n^{2/3} s_t(0))\\[1ex]
            -q_{\alpha}(n^{2/3} s_t(0)) & -u_{\alpha}(n^{2/3} s_t(0))
        \end{pmatrix}
        e^{-i(\pi n\omega_t+\alpha\theta)\sigma_3} (\widehat P_+^{(\infty)})^{-1}(0)
        D_\infty^{-\sigma_3}.
\end{multline}
Combining (\ref{definition: st}), (\ref{definition: f}), and (\ref{psitassqrt})
we see that $n^{2/3} s_t(0) = s_{t,n}$ as defined in (\ref{definition: s recursie}).
From (\ref{defhatPinfty}), (\ref{Pinfty}), and (\ref{definition: beta})
it follows that
\[ \widehat P_+^{(\infty)}(0) =
    \begin{pmatrix} \frac{\beta_+(0) + \beta_+(0)^{-1}}{2}
        & \frac{\beta_+(0) - \beta_+(0)^{-1}}{2i} \\
        \frac{\beta_+(0) - \beta_+(0)^{-1}}{2i} &
        \frac{\beta_+(0) + \beta_+(0)^{-1}}{2} \end{pmatrix}, \]
where $\beta_+(0) = e^{i \pi/4} \left(-b_t/a_t\right)^{1/4}$.

We insert this into (\ref{resDelta1}) and after some straightforward
calculations we find
\begin{equation}\label{resDelta2}
- \textrm{Res}(\Delta^{(1)};0) = D_{\infty}^{\sigma_3}
    \left( r_1\sigma_1 + r_2\sigma_2+r_3\sigma_3 \right) D_{\infty}^{-\sigma_3},
\end{equation}
where the Pauli matrices are $\sigma_1= \left(\begin{smallmatrix}0&1\\1&0\end{smallmatrix}\right)$,
$\sigma_2=\left(\begin{smallmatrix} 0&-i\\i&0 \end{smallmatrix}\right)$, and
$\sigma_3=\left(\begin{smallmatrix} 1&0\\0&-1 \end{smallmatrix}\right)$,
and
\begin{align} \nonumber
    r_1
        & = -\frac{1}{2ic}\left(u_{\alpha}(s_{t,n}) \frac{b_t-a_t}{2\sqrt{-a_tb_t}}
            + q_{\alpha}(s_{t,n}) \frac{b_t+a_t}{2\sqrt{-a_tb_t}}
            \sin (2\pi n \omega_t + 2\alpha\theta_t) \right)
    \\[1ex]
        & = \label{X1}
            -\frac{1}{2ic}\left(u_{\alpha}(s_{t,n}) \frac{b-a}{2\sqrt{-ab}}
            + q_{\alpha}(s_{t,n}) \frac{b+a}{2\sqrt{-a b}}
            \sin (2\pi n \omega_t + 2\alpha\theta) \right) + \bigO(n^{-2/3}),
\end{align}
\begin{align} \nonumber
    r_2
        & =\frac{q_{\alpha}(s_{t,n})}{2c}\cos(2\pi n \omega_t + 2\alpha \theta_t)
    \\[1ex]
        & = \frac{q_{\alpha}(s_{t,n})}{2c}\cos(2\pi n \omega_t + 2\alpha \theta)
            + \bigO(n^{-2/3}), \label{X2}
\end{align}
\begin{align}
        \nonumber
    r_3 & =\frac{1}{2c}\left(q_{\alpha}(s_{t,n}) \frac{b_t-a_t}{2\sqrt{-a_tb_t}}
    \sin(2\pi n \omega_t + 2\alpha \theta_t) +
      u_{\alpha}(s_{t,n}) \frac{b_t+a_t}{2\sqrt{-a_tb_t}}  \right)
    \\[1ex]
      & = \label{X3}
      \frac{1}{2c}\left(q_{\alpha}(s_{t,n}) \frac{b-a}{2\sqrt{-ab}}
    \sin(2\pi n \omega_t + 2\alpha \theta) +
      u_{\alpha}(s_{t,n}) \frac{b+a}{2\sqrt{-ab}}  \right) + \bigO(n^{-2/3}),
\end{align}
where we used (\ref{abanalytic}).

From (\ref{vRn}) it follows that
\begin{equation}\label{Rn}
R(z)=I+\frac{R^{(1)}(z)}{n^{1/3}}+\bigO(n^{-2/3}),
\end{equation}
where $R^{(1)}_+ = R^{(1)}_- + \Delta^{(1)}$ on $\partial U_{\delta}$ and
$R^{(1)}(z)\to 0$ as $z\to\infty$. Since $\Delta^{(1)}$ is analytic with a
simple pole at $z=0$, we can find explicitly
\begin{equation}\label{R1n}
    R^{(1)}(z)=
        \begin{cases}
            -\frac{1}{z}\textrm{Res}(\Delta^{(1)};0)+\Delta^{(1)}(z),
            &\mbox{for $z\in U_\delta$,}
        \\[1ex]
            -\frac{1}{z}\textrm{Res}(\Delta^{(1)};0),
            &\mbox{for $z\in\mathbb{C}\setminus \overline{U}_\delta$.}
        \end{cases}
\end{equation}
As in \cite{DKMVZ1} the matrix valued function $R$ has the
following asymptotic behavior at infinity,
\begin{equation}\label{Rinfty}
    R(z)=I+\frac{R_1}{z}+\frac{R_2}{z^2}+\bigO(1/z^3),
        \qquad\textrm{as } z\to\infty.
\end{equation}
The compatibility with (\ref{Rn}) and (\ref{R1n}) yields that
\begin{equation} \label{R1form}
R_1=-\textrm{Res}(\Delta^{(1)};0) n^{-1/3} +\bigO(n^{-2/3}), \qquad R_2=\bigO(n^{-2/3}).
\end{equation}

Now, we are ready to determine the asymptotics of the recurrence
coefficients.
\begin{varproof}\textbf{of Theorem \ref{theorem: recursie}.}
Note that by (\ref{Yinfty}), (\ref{Pinfty-infty}),
(\ref{ginfty}) and (\ref{Rinfty}),
\begin{equation}\label{definition: Y1}
    Y_1= e^{-\frac{1}{2}n\ell_t\sigma_3}\left[P_1^{(\infty)}+G_1+R_1\right]e^{\frac{1}{2}n\ell_t\sigma_3}
\end{equation}
and
\begin{equation}\label{definition: Y2}
    Y_2= e^{-\frac{1}{2}n\ell_t\sigma_3}\left[P_2^{(\infty)}+G_2+R_2
        +R_1 P_1^{(\infty)}+\left(P_1^{(\infty)}+R_1\right)G_1\right]e^{\frac{1}{2}n\ell_t\sigma_3}
\end{equation}

    We start with the recurrence coefficient $a_{n,N}$. Inserting
    (\ref{definition: Y1}) into (\ref{recurrence in Y}) and using (\ref{definition: P1infty})
    and the facts that $(G_1)_{12}=(G_1)_{21}=0$ (by (\ref{G1def})),
    and $(R_1)_{12}(R_1)_{21}=\bigO(n^{-2/3})$ (by (\ref{R1form})),
    we obtain
    \begin{align*}
        a_{n,N} &=
            \left[(P_1^{(\infty)})_{12} (P_1^{(\infty)})_{21} +
                (P_1^{(\infty)})_{12} (R_1)_{21}+(P_1^{(\infty)})_{21}(R_1)_{12}+\bigO(n^{-2/3})\right]^{1/2}
            \\[1ex]
            &=\left[\left(\frac{b-a}{4}\right)^2+i\frac{b-a}{4}\left(D_\infty^2 (R_1)_{21}
                -D_\infty^{-2}(R_1)_{12}\right)+\bigO(n^{-2/3})\right]^{1/2}
            \\[1ex]
            &=\frac{b-a}{4}+\frac{i}{2}\left(D_\infty^2 (R_1)_{21}
                -D_\infty^{-2}(R_1)_{12}\right)+\bigO(n^{-2/3}).
    \end{align*}
    From (\ref{R1form}) and (\ref{resDelta2}) we then arrive at,
    \begin{eqnarray}
        a_{n,N}&=&\frac{b-a}{4}-r_2 n^{-1/3}+\bigO(n^{-2/3})\nonumber\\
        &=& \frac{b-a}{4}-\frac{q_{\alpha}(s_{t,n})\cos(2\pi n \omega_t + 2 \alpha \theta)}{2c}
        n^{-1/3}+\bigO(n^{-2/3}). \label{anNfinal}
    \end{eqnarray}

    Next, we consider the recurrence coefficient $b_{n,N}$.
    Inserting (\ref{definition: Y1}) and (\ref{definition: Y2}) into
    (\ref{recurrence in Y}), and using the facts that
    $(G_1)_{11}+(G_1)_{22}=0$ (by (\ref{G1def})), and $(R_2)_{12}=\bigO(n^{-2/3})$
    (by (\ref{R1form})) we obtain
    \begin{align*}
        b_{n,N} &= (P_1^{(\infty)})_{11}+(R_1)_{11}+
            \frac{(P_2^{(\infty)})_{12}+
            (R_1
            P_1^{(\infty)})_{12}+\bigO(n^{2/3})}{(P_1^{(\infty)}+R_1)_{12}}
        \\[1ex]
        &= (P_1^{(\infty)})_{11}+ (R_1)_{11}+
        \left(1-\frac{(R_1)_{12}}{(P_1^{(\infty)})_{12}}+\bigO(n^{-2/3})\right)
        \\[1ex]
        &\qquad\qquad\qquad\qquad\times\,
        \left(\frac{(P_2^{(\infty)})_{12}}{(P_1^{(\infty)})_{12}}+(R_1)_{11}+
        \frac{(P_1^{(\infty)})_{22}}{(P_1^{(\infty)})_{12}}(R_1)_{12}+\bigO(n^{-2/3})\right).
    \end{align*}
    From equations (\ref{definition: P1infty}), (\ref{definition: P2infty}),
    (\ref{R1form}), and (\ref{resDelta2}), we then arrive at
   \begin{align}
        \nonumber
        b_{n,N}
        &=
        \frac{b+a}{2}+2(R_1)_{11}+2i\frac{b+a}{b-a}D_\infty^{-2}
        (R_1)_{12}+\bigO(n^{-2/3}) \\[1ex]
        &=\frac{b+a}{2}+2\left(r_3+i\frac{b+a}{b-a}(r_1-ir_2)\right)
            + \bigO(n^{-2/3}). \label{bnNformula2}
    \end{align}

Using (\ref{X1}), (\ref{X2}), and (\ref{X3}) in (\ref{bnNformula2}) we
will see that the terms containing $u_{\alpha}$ cancel against each other.
What remains are the terms containing $q_{\alpha}$:
\begin{multline}
    b_{n,N} =
            \frac{b+a}{2}
            + \frac{q_{\alpha}(s_{t,n})}{c}
            \left[
                \frac{b+a}{b-a} \cos (2\pi n \omega_t+2\alpha \theta)
                + \frac{2\sqrt{-ab}}{b-a} \sin(2\pi n \omega_t + 2\alpha \theta)
            \right] n^{-1/3}
        \\[1ex]
            + \bigO(n^{-2/3}).
\end{multline}
Since $\frac{b+a}{b-a} = \sin \theta$ and
$\frac{2\sqrt{-ab}}{b-a} = \cos \theta$, we can combine the two
terms within square brackets and the result is
\begin{equation} \label{bnNfinal}
    b_{n,N} = \frac{b+a}{2}
    + \frac{q_{\alpha}(s_{t,n})\sin(2\pi n \omega_t + (2\alpha +1) \theta)}{c}
      n^{-1/3}+\bigO(n^{-2/3}).
    \end{equation}
Theorem \ref{recursie} is proven by (\ref{anNfinal}) and (\ref{bnNfinal}).
\end{varproof}

\section*{Acknowledgments}

We are grateful to Pavel Bleher and Alexander Its for very useful
and stimulating discussions.

The authors are supported by FWO research projects G.0176.02 and
G.0455.04. The second author is also supported by K.U.Leuven
research grant OT/04/24, by INTAS Research Network NeCCA
03-51-6637, by NATO Collaborative Linkage Grant PST.CLG.979738, by
grant BFM2001-3878-C02-02 of the Ministry of Science and
Technology of Spain and by the European Science Foundation Program
Methods of Integrable Systems, Geometry, Applied Mathematics
(MISGAM) and the European Network in Geometry, Mathematical
Physics and Applications (ENIGMA). The third author is
Postdoctoral Fellow of the Fund for Scientific Research - Flanders
(Belgium). He is also grateful to the Department of Mathematics of
the Ruhr Universit\"at Bochum where he has spent the academic year
2004-2005, for hospitality.

\bigskip\noindent
{\sc Department of Mathematics, Katholieke Universiteit Leuven,\\
Celestijnenlaan 200B, 3001 Leuven, Belgium} \newline E-mail
adresses: \newline tom.claeys@wis.kuleuven.be \newline
arno.kuijlaars@wis.kuleuven.be \newline
maarten.vanlessen@wis.kuleuven.be

\end{document}